\documentclass[aps,prl,twocolumn,floatfix,longbibliography,twocolumn, superscriptaddress]{revtex4-1}

\usepackage[T2A,T1]{fontenc}
\usepackage{amsmath}
\usepackage{amssymb}
\usepackage{graphicx}
\usepackage{natbib}
\usepackage[dvipsnames]{xcolor}
\usepackage[utf8]{inputenc}

\usepackage{soul}
\usepackage{booktabs}
\usepackage{tabularx}

\newcommand\ket[2][]{#1\lvert {#2} #1\rangle}

\begin{document}%

\title{Optimizing magnetic coupling in lumped element superconducting resonators for molecular spin qubits}

\author{Marcos Rub\'{i}n-Osanz}
\affiliation{Instituto de Nanociencia y Materiales de Arag\'{o}n (INMA), CSIC-Universidad de Zaragoza, Zaragoza 50009, Spain}

\author{David Rodriguez}
\affiliation{Centro de Astrobiolog\'{i}a (CSIC-INTA), Torrej\'on de Ardoz, 28850 Madrid, Spain}

\author{Ignacio Gimeno}
\affiliation{Instituto de Nanociencia y Materiales de Arag\'{o}n (INMA), CSIC-Universidad de Zaragoza, Zaragoza 50009, Spain}

\author{Wenzel Kersten}
\affiliation{Vienna Center for Quantum Science and Technology, Atominstitut, TU Wien, A-1020 Vienna, Austria}

\author{Nerea Gonz\'alez-Pato}
\affiliation{Institut de Ci\`{e}ncia de Materials de Barcelona (ICMAB-CSIC), Networking Research Center on Bioengineering Biomaterials and Nanomedicine (CIBER-BBN), Campus de la UAB, Bellaterra, 08193 Spain}

\author{Mar\'{i}a C. Pallar\'{e}s}
\affiliation{Instituto de Nanociencia y Materiales de Arag\'{o}n (INMA), CSIC-Universidad de Zaragoza, Zaragoza 50009, Spain}
\affiliation{Laboratorio de Microscop\'{i}as Avanzadas (LMA), Universidad de Zaragoza, Zaragoza 50018, Spain}

\author{Sebasti\'{a}n Roca-Jerat}
\affiliation{Instituto de Nanociencia y Materiales de Arag\'{o}n (INMA), CSIC-Universidad de Zaragoza, Zaragoza 50009, Spain}

\author{Marina C. de Ory}
\affiliation{Centro de Astrobiolog\'{i}a (CSIC-INTA), Torrej\'on de Ardoz, 28850 Madrid, Spain}

\author{Marta Mas-Torrent}
\affiliation{Institut de Ci\`{e}ncia de Materials de Barcelona (ICMAB-CSIC), Networking Research Center on Bioengineering Biomaterials and Nanomedicine (CIBER-BBN), Campus de la UAB, Bellaterra, 08193 Spain}

\author{J. Alejandro de Sousa}
\affiliation{Institut de Ci\`{e}ncia de Materials de Barcelona (ICMAB-CSIC), Networking Research Center on Bioengineering Biomaterials and Nanomedicine (CIBER-BBN), Campus de la UAB, Bellaterra, 08193 Spain}

\author{Lorenzo Tesi}
\affiliation{Institute of Physical Chemistry, University of Stuttgart, 70569 Stuttgart, Germany}
\affiliation{Center for Integrated Quantum Science and Technology, University of Stuttgart, 70569 Stuttgart, Germany}

\author{Daniel Granados}
\affiliation{IMDEA Nanociencia, Cantoblanco, 28049 Madrid, Spain}

\author{Jaume Veciana}
\affiliation{Institut de Ci\`{e}ncia de Materials de Barcelona (ICMAB-CSIC), Networking Research Center on Bioengineering Biomaterials and Nanomedicine (CIBER-BBN), Campus de la UAB, Bellaterra, 08193 Spain}

\author{David Zueco}
\affiliation{Instituto de Nanociencia y Materiales de Arag\'{o}n (INMA), CSIC-Universidad de Zaragoza, Zaragoza 50009, Spain}

\author{Anabel Lostao}
\affiliation{Instituto de Nanociencia y Materiales de Arag\'{o}n (INMA), CSIC-Universidad de Zaragoza, Zaragoza 50009, Spain}
\affiliation{Laboratorio de Microscop\'{i}as Avanzadas (LMA), Universidad de Zaragoza, Zaragoza 50018, Spain}
\affiliation{Fundaci\'on ARAID, Zaragoza 50018, Spain}

\author{J\"org Schmiedmayer}
\affiliation{Vienna Center for Quantum Science and Technology, Atominstitut, TU Wien, A-1020 Vienna, Austria}

\author{Imma Ratera}
\affiliation{Institut de Ci\`{e}ncia de Materials de Barcelona (ICMAB-CSIC), Networking Research Center on Bioengineering Biomaterials and Nanomedicine (CIBER-BBN), Campus de la UAB, Bellaterra, 08193 Spain}

\author{Joris van Slageren}
\affiliation{Institute of Physical Chemistry, University of Stuttgart, 70569 Stuttgart, Germany}
\affiliation{Center for Integrated Quantum Science and Technology, University of Stuttgart, 70569 Stuttgart, Germany}

\author{N\'uria Crivillers}
\affiliation{Institut de Ci\`{e}ncia de Materials de Barcelona (ICMAB-CSIC), Networking Research Center on Bioengineering Biomaterials and Nanomedicine (CIBER-BBN), Campus de la UAB, Bellaterra, 08193 Spain}

\author{Alicia Gomez}
\affiliation{Centro de Astrobiolog\'{i}a (CSIC-INTA), Torrej\'on de Ardoz, 28850 Madrid, Spain}

\author{Fernando Luis}
\email{fluis@unizar.es}
\affiliation{Instituto de Nanociencia y Materiales de Arag\'{o}n (INMA), CSIC-Universidad de Zaragoza, Zaragoza 50009, Spain}

\begin{abstract}
We engineer lumped-element superconducting resonators 
that maximize magnetic coupling to molecular spin 
qubits, achieving record single-spin couplings up to 
$100$ kHz and collective couplings exceeding $10$ MHz.
The resonators interact with PTMr 
organic free radicals, model spin systems with 
$S=1/2$ and a quasi-isotropic $g \simeq 2$, dispersed 
in polymer matrices. The highest collective spin-
photon coupling strengths are 
attained with resonators having large inductors, 
which therefore interact with most spins in the 
molecular ensemble. By contrast, the 
coupling of each individual spin $G_{1}$ is 
maximized in resonators having a minimum size 
inductor, made of a single wire. The same 
platform has been used to study spin 
relaxation and spin coherent dynamics in the 
dispersive regime, when spins are energetically 
detuned from the resonator. We 
find evidences for the Purcell effect, i.e. the 
photon induced relaxation of those spins that are 
most strongly coupled to the circuit. The rate of 
this process gives access to the  
distribution of single spin photon couplings in a 
given device. For resonators with a $50$ nm wide 
constriction at the inductor center, single 
maximum $G_{1}$ values reach $\sim 100$ kHz. Pumping 
the spins with strong pulses fed through an 
independent transmission line induces coherent Rabi 
oscillations. The spin excitation then proceeds via either 
direct resonant processes induced by the main pulse 
frequency or, in the case of square-shaped pulses, 
via the excitation of the cavity by sideband frequency 
components. The latter process measures the cavity 
mode hybridization with the spins and can be 
eliminated by using Gaussian shaped pulses. These 
results establish a scalable route toward integrated 
molecular-spin quantum processors.
\end{abstract}
\maketitle
%%%%%%%%%%%%%%%%%%%%%%%%%%%%%%%%%%% 
%ABS 
%%%%%%%%%%%%%%%%%%%%%%%%%%%%%%%%%% 
%%%%%%%%%%%%%%%%%%%%%%%%%%%%%%%
%%%% INTRO 
%%%%%%%%%%%%%%%%%%%%%%%%%%%%%%%

\section{Introduction}
\label{sec:intro}

Magnetic molecules are seen as potentially advantageous 
realization of spin qubits, due to their easily 
controllable purity and reproducibility, and to the fact 
that the qubit properties can be tuned by 
chemistry \cite{GaitaArino2019,Atzori2019,Wasiliewski2020,Chiesa2024}.
The ability to control the molecular composition and 
environment, as well as the local coordination of the 
spin centers has been exploited to isolate spins from 
magnetic noise sources \cite{Wedge2012,Bader2014,Zadrozny2015a,Shiddiq2016}.
This approach has led to systems showing spin 
coherence times exceeding tens of micro-seconds at 
low temperatures, some of them retaining coherence 
even up to room temperature \cite{Bader2014,Atzori2016}. Besides, it also gives 
opportunities for scaling up computational resources, 
e.g. by encoding multiple qubits within each 
molecule \cite{Luis2011,Aromi2012,Aguila2014,Fernandez2016,Ferrando-Soria2016,Jenkins2017,Godfrin2017,Hussain2018,Moreno-Pineda2018,Luis2020,Macaluso2020,Carretta2021,Chiesa2024, Roca2025simulating}. 
Of especial 
interest is the ability of using the additional 
redundancy to perform quantum simulations
\cite{Chiesa2023,Chizzini2024,Chicco2024,Roca2025simulating} or for 
embedding quantum error correction 
codes 
\cite{Hussain2018,Macaluso2020,Chiesa2020,Chiesa2022,Chiesa2022b,Lim2023,Mezzadri2024,Lim2025}. 
The fact that these properties are defined at the molecular scale and that molecules 
can be integrated in different environments from solution \cite{Domingo20212,Malavolti2018,Serrano2020} are also key 
ingredients for their potential implementation in scalable devices. However, wiring up such molecules into a quantum 
computing architecture, able to control and read out their spin states, remains very challenging. 

Recent proposals \cite{Jenkins2013,Jenkins2016,Chiesa2023} aim to 
adapt techniques and protocols from circuit quantum 
electrodynamics, originally developed for 
superconducting quantum processors \cite{Blais2004, Wallraff2004, Majer2007,Schoelkopf2008,Blais2021} to 
the realm of magnetic molecules. They rely on 
coherently coupling single spins to superconducting resonators 
and transmission lines to induce transitions between 
different spin states \cite{Castro2022,Hernandez2025}, 
perform a non-demolition readout of the spin state in 
the dispersive regime \cite{Gomez-Leon2022}, and to 
introduce the effective interactions between remote 
molecules \cite{Carretta2021,Gomez-Leon2022b} that 
are key to achieve full scalability.

Early theoretical schemes \cite{Jenkins2013,Jenkins2016} were based on coplanar 
waveguide resonators \cite{Goppl2008}. Experiments performed with 
these devices have shown the ability to reach a strong, 
or coherent, coupling of relatively large molecular spin ensembles to single photon 
excitations \cite{Ghirri2015,Ghirri2016,Bonizzoni2017,Mergenthaler2017}.
However, these circuits have significant limitations in 
design, e.g. in order to match the line impedance, 
which constrain the maximum attainable strength of the microwave magnetic field 
generated by the resonator. This effectively precludes attaining a sufficiently high 
interaction to each individual spin. 

In this work, we instead focus on lumped element 
resonators (LERs), made of inductor-capacitor circuits, 
which offer a promising alternative to overcome these 
limitations. Its two components ($L$ and $C$) can be 
designed independently of each other. This enables 
magnetic-field localization by geometric design, 
which is central to optimizing spin–photon 
coupling \cite{Bienfait2016,Eichler2017,Probst2017,Rollano2022,Rubin2024}. Besides, 
they are easier to combine with one or multiple transmission lines that generate 
external driving fields to control the spins and to read out the LER state. 
And last, but not least, several of them can be parallel coupled to 
a common transmission line, which allows multiplexing the readout and facilitates 
scaling up. Recent simulations show that these devices provide a suitable platform 
for the development of a scalable hybrid quantum processor based on molecular 
spins \cite{Chiesa2023}.

%%%%%%%%%%%%% fig 1
%
\begin{figure}[h!]
\begin{centering}
\includegraphics[width=0.9\columnwidth, angle=-0]{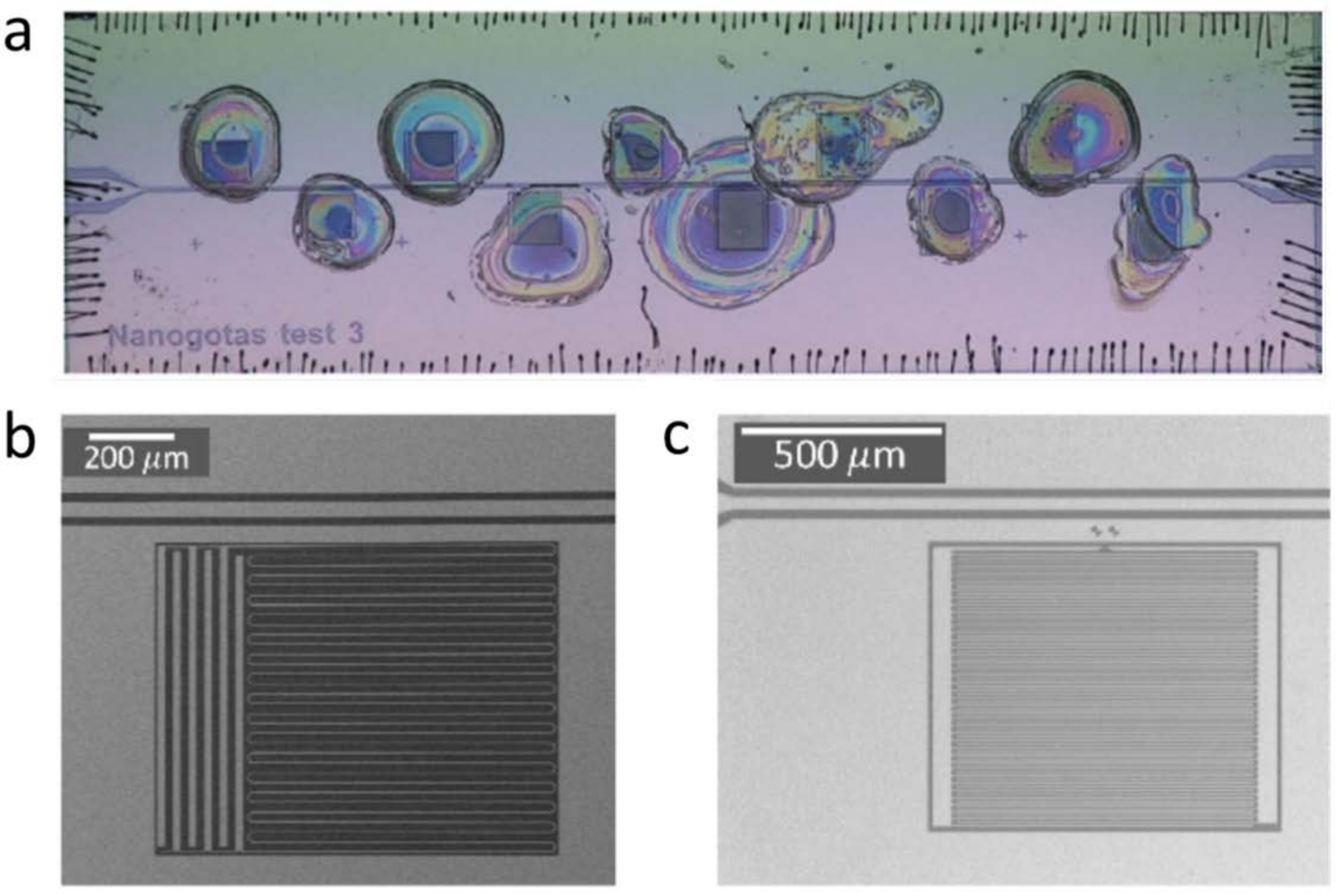}
\caption{(a) Optical microscopy image of a chip hosting ten low-impedance Nb LERs 
with different resonance frequencies $1.5$ GHz$\leq \omega_{\rm r}/ 2 \pi \leq 3$ GHz, 
all coupled to the same readout transmission 
line and hosting dry deposits of free radical PTMr molecules embedded in 
polystyrene. (b) Image of a $1.971$~GHz LER with a large meandering inductor, 
designed to optimally couple to large spin sample volumes. (c) Image of a 
$1.767$~GHz low-impedance LER with a small-size inductor (a $12$~$\mu m$ wide wire 
near the readout line) tailored to optimally couple to small sample volumes.}
\label{fig:LERs}
\end{centering}
\end{figure}

We explore the coupling to LERs of ensembles of the 
simplest molecular spin qubits, perchlorotriphenylmethyl 
(PTMr) organic free radicals hosting an unpaired 
electron with $S=1/2$ and $g \simeq 2$, which are 
stable, have a well-defined nearly isotropic 
resonance line, can show quite long spin coherence 
times in the solid state and be 
optimally integrated into the chips \cite{Dai2018,Schafter2023}. The main 
objectives are to tune and optimize, via 
circuit design, the coupling to spin ensembles but 
also, and especially, at the single molecule level, 
to best visualize and determine such coupling and to 
illustrate experimentally the implementation of the 
basic ingredients of a hybrid quantum platform. 
Section \ref{sec:experimental} describes the design, 
simulation and fabrication of the superconducting 
circuits, the integration of the 
molecular spin samples and the microwave measurement 
set-ups. Section \ref{sec:expresults} discusses the 
main results, starting from the spectroscopic 
characterization of the molecular spins by magnetic 
resonance and then focusing on the results of either 
continuous wave or time-resolved microwave 
transmission experiments. The last section 
\ref{sec:conclusions} summarizes the main conclusions 
of this work.

%%%%%%%%%%%%% fig 2
%
\begin{figure}[h!]
\begin{centering}
\includegraphics[width=1.0\columnwidth, angle=-0]{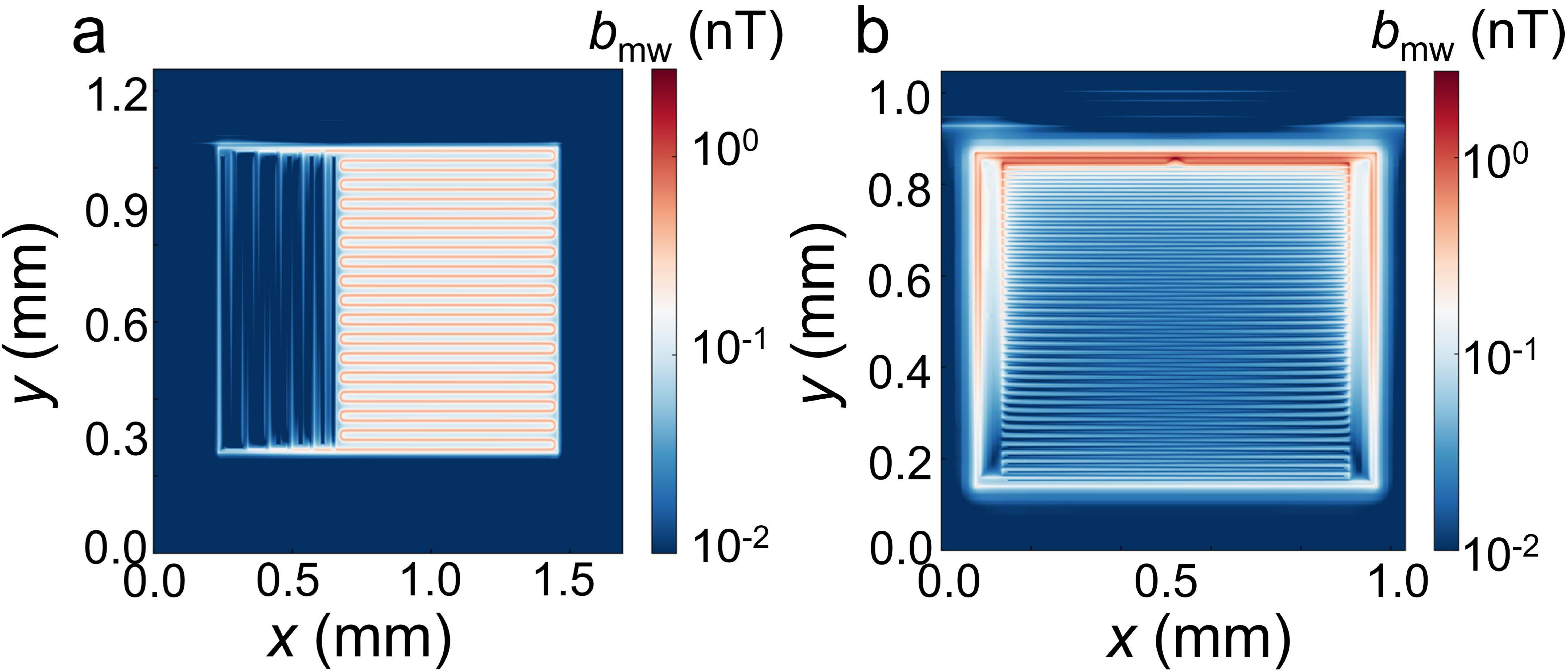}
\caption{Simulation of the microwave magnetic field amplitude $b_{\rm mw}$ generated by a LER vacuum fluctuations $1~\mu \mathrm{m}$ above the chip surface. Two LER 
designs are considered, with (a) a meandered inductor $L_{\rm HL}$ and (b) a single 
wire inductor $L_{\rm LL}$ with a central constriction. Both simulations were 
performed for the same photon energy ($\omega_{\rm r} / 2 \pi = 1.714$ GHz). 
The ratio between the maximum microwave currents of each design is approximately 
$0.22$, in agreement with $(L_{\rm LL} / L_{\rm HL})^{1/2} \approx 0.17$.}
\label{fig:field intensity LERs}
\end{centering}
\end{figure}

\section{Experimental details}
\label{sec:experimental}
\subsection{Design and fabrication of superconducting lumped-element resonators}
\label{ssec:LERs}

Each of the chips studied in this work contains 
several LERs with different resonance frequencies $\omega_{\rm r}$, 
side-coupled to the same coplanar transmission line, as shown in Fig. 1. The 
patterns are lithographically etched on a $150$ nm thick Nb or NbTiN layer 
on a silicon substrate \cite{Rollano2022,Rubin2024}. LERs made of NbTiN show a better 
stability of the resonance against magnetic field. 

%%%%%%%%%%%%% fig 3
\begin{figure*}
\begin{centering}
\includegraphics[width=1.98\columnwidth, angle=-0]{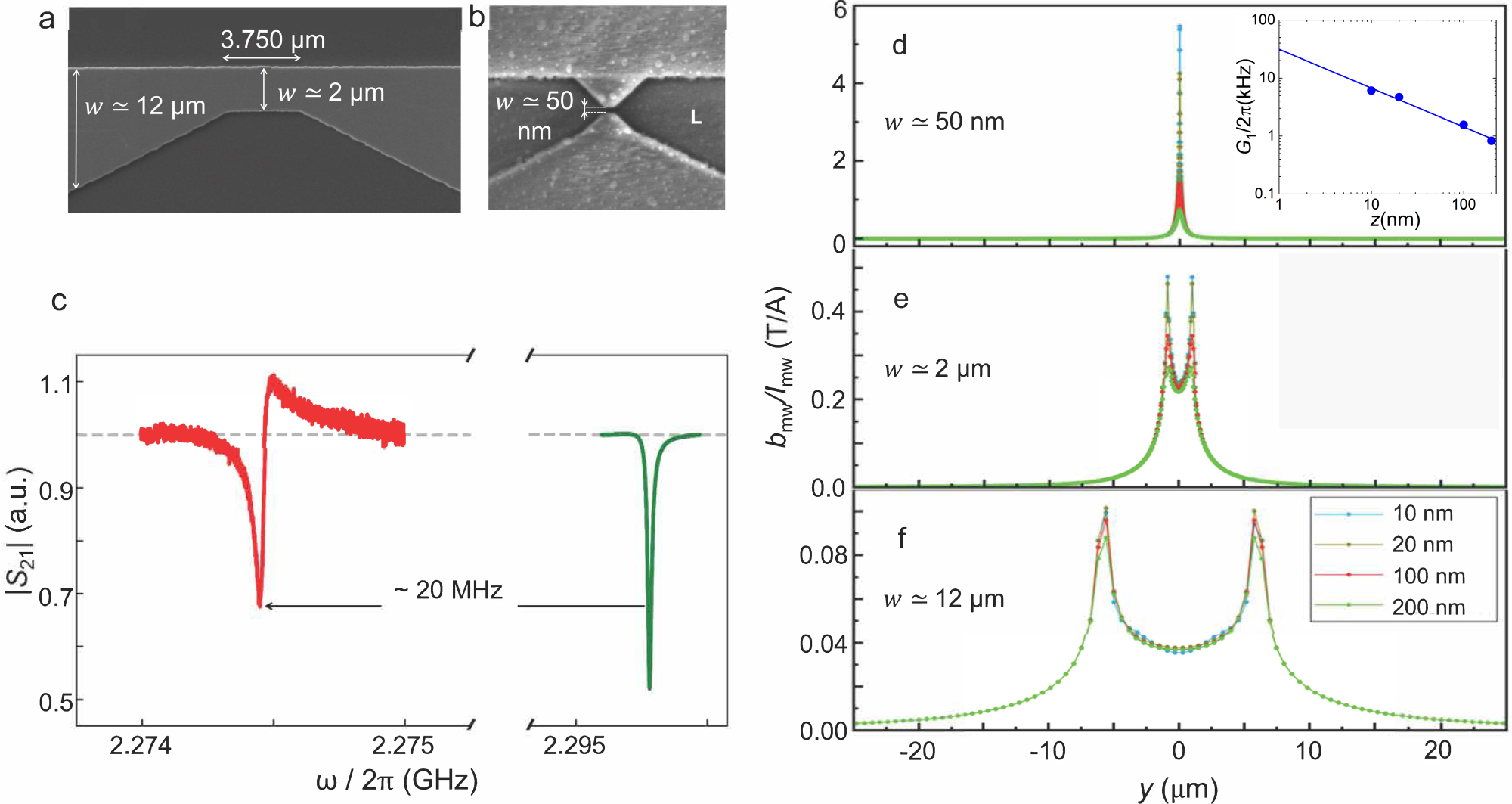}
\caption{SEM images of the inductor center of a low-impedance Nb LER (a) 
before and (b) after the fabrication of a $50$~nm wide nanoconstriction. The 
white spots in (b) are nanoscopic drops of PTMr 
embedded in a polymer matrix. (c) Microwave 
transmission through this chip measured at $T=12$ mK 
and at zero magnetic field near the LER resonance 
frequency before (green) and after (red) the 
fabrication of the nanoconstriction. (d, e, f) Finite element simulations of the magnetic 
field generated by single-wire 
inductors of different widths $w$, corresponding to the different regions of the LER inductors shown 
in panels (a) and (b) (d: $w = 50$~nm, e: $w = 2$~$\mu$m, f: $w = 12$~$\mu$m). They were  
calculated at various heights $z=10$~nm, $20$~nm, $100$~nm and $200$~nm above the 
chip surface. The magnetic fields are normalized to 
the current in order to account 
for differing inductances, while keeping the photon 
energy constant ($\omega_{\rm r} / 2 \pi = 1.71 \, \mathrm{GHz}$). The inductance values are as 
follows: $0.429$~nH for the wire with a $50$~nm 
nanoconstriction, $0.420$~nH for the wire with 
a $2$~$\mu$m constriction, and $0.231$~nH 
for the $12$~$\mu$m wire, with corresponding 
current values of $39.39$~nA, $36.78$~nA, 
and $40.96$~nA, respectively. The inset of (d) shows a log-log plot of the coupling $G_{1}$ to a single 
$S=1/2$ spin located at a height $z$ over the center ($y=0$) of a $50$ nm wide inductor. The line is a 
least squares linear fit extrapolating to $G_{1} \sim 32$ kHz for $z = 1$ nm.}
\label{fig:frec before after fib and fields}
\end{centering}
\end{figure*}

The LER inductor geometry defines 
the main parameters that ultimately determine the coupling to 
spins, such as $\omega_{\rm r}$, which must be within 
the ranges attainable for the spin resonance 
frequencies $\Omega_{S}$ at suitably low 
magnetic fields, the coupling $\kappa_{\rm c}$ to the 
readout line and the orientation 
and spatial distribution of the microwave magnetic 
field $\vec{b}_{\rm mw}$. We use 
electromagnetic simulations based on the Sonnet 
package \cite{Rautio1997} in order to 
tailor $\omega_{\rm r}$ and the current distribution 
of each LER. In this work, two characteristic LER 
designs were used in order to maximize either the 
mode volume or the microwave field intensity: high-
inductance LERs, with meander-shaped 
inductors as that shown in Fig.~\ref{fig:LERs}b, and 
low-inductance LERs (Fig.~\ref{fig:LERs}c), whose 
inductor was reduced to a single straight wire. The 
capacitor was adapted to maintain $\omega_{\rm r}$ within comparable ranges for 
both designs. The resonance frequencies lie between 
$1.5$ and $2$ GHz for the former 
and between $1.5$ and $3$~GHz for the latter.

Figure~\ref{fig:field intensity LERs} shows $b_{\rm mw}$
calculated at a distance of $1$~$\mu \mathrm{m}$ 
above the surface of two LERs having 
the same $\omega_{\rm r}$, thus the same photon 
energy but with either high $L_{\rm HL}$ or low 
$L_{\rm LL}$ inductances. Near resonance, half the 
energy of each photon is stored in the microwave 
current $I_{\rm mw}$ at the inductor, meaning that 
$(1/2) \hbar \omega_{\rm r} \simeq (1/2) LI_{\rm mw}^{2}$. Therefore, the maximum 
current, thus also the maximum $b_{\rm mw}$, 
approximately scale with $L^{-1/2}$, as 
the simulated field profiles of
Fig. \ref{fig:field intensity LERs} confirm. These 
results show that low-inductance LERs are best to 
optimize the coupling to each individual spin within 
the small region defined by the inductor line \cite{Eichler2017}. 

%
%%%%%%%%%%%%% fig 5
\begin{figure*}
\begin{centering}
\includegraphics[width=1.98\columnwidth, angle=-0]{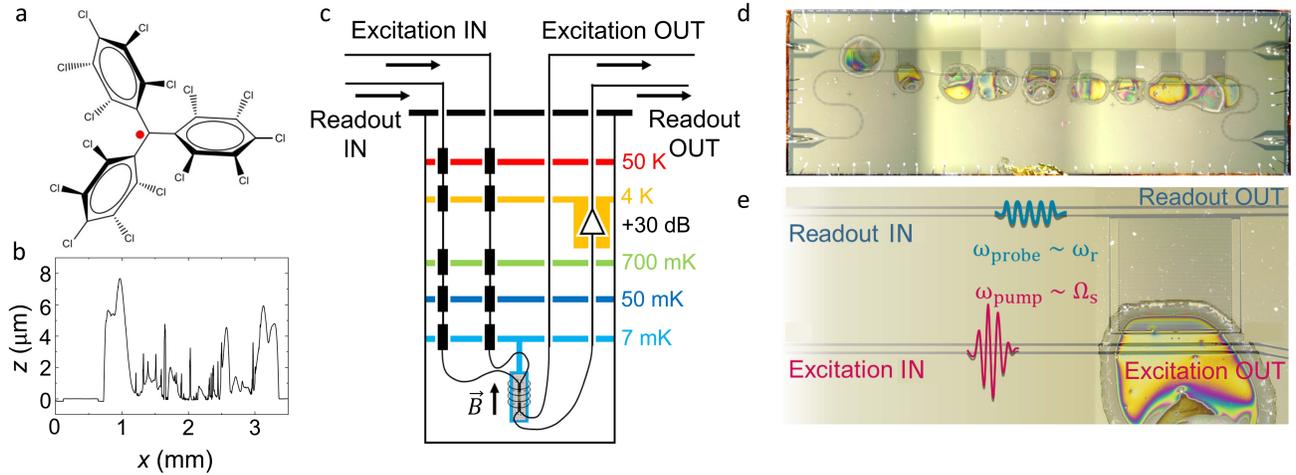}
\caption{(a) Molecular structure of PTMr, showing the free 
electron on the central $\alpha$ carbon atom. (b) Topographic 
atomic force microscopy height profiles measured on a dry PTMr/PS drop 
deposited onto one of the Nb LERs shown in 
Fig. \ref{fig:LERs}. (c) Scheme of the set-up for 
transmission and pump-probe experiments. The cryostat, 
either an $^{3}$He-$^{4}$He dilution or an adiabatic 
demagnetization refrigerator, has several cryogenic 
coaxial cables that drive the different input and output 
microwave signals to and from the chip, respectively. Input 
lines incorporate either 0 dB or -10 dB attenuators 
at each constant temperature plate. Output readout lines 
are amplified at $T = 4.2$ K 
before reaching the digital readout electronics while 
excitation lines, which drive higher power pulses, are 
directly fed into a digital high frequency oscilloscope. 
(d) Chip hosting multiple low-inductance NbTiN LERs coupled 
to independent readout and control transmission lines. The 
latter, inductively coupled to the PTMr molecular deposits, 
can be used to induce spin excitations through strong 
microwave pulses whereas the 
former allows reading out the LER resonance. (e) Zoom of one of the 
LERs in the same chip.}
\label{fig:setup}
\end{centering}
\end{figure*}

In addition to choosing the proper LER design, fine tuning 
the inductor dimensions can also help in further enhancing 
$b_{\rm mw}$. In particular, reducing 
the inductor width $w$ leads to a local increase in the 
superconducting current density \cite{Jenkins2014} and, 
therefore, it also increases the maximum $b_{\rm mw}$. 
However, this also leads to a larger inductance. In order 
to limit the latter, a constriction can be fabricated at a 
small region near the center of the inductor line 
(see Fig.~\ref{fig:frec before after fib and fields}a), which 
effectively defines a smaller mode 
volume with an enhanced $b_{\rm mw}$, as the simulations 
show (Figs. \ref{fig:frec before after fib and fields}d,\ref{fig:frec before after fib and fields}e and 
\ref{fig:frec before after fib and fields}f). Nanoscopic 
constrictions, typically 
$50$ nm wide and $500$ nm long such as that shown in Fig. 
\ref{fig:frec before after fib and fields}b, were made by milling down 
the inductor line with a focused Ga$^{+}$ ion 
beam \cite{Jenkins2014,Gimeno2020}. Figure~\ref{fig:frec before after fib and fields} shows the 
microwave transmission measured near the 
resonance frequency of this LER before 
and after the fabrication of the nanoconstriction. A $\sim 20$ MHz downward shift in 
$\omega_{\rm r}/2 \pi$ is observed in all the modified LERs, as expected from 
the slight inductance increase caused by the reduction of 
the wire width. The inductance increase estimated from this shift is between $0.02$ 
and $0.09 \, \mathrm{nH}$. Besides, the internal quality factor 
$Q_{\rm i} \equiv \omega_{\rm r}/\kappa_{\rm i}$, with $\kappa_{\rm i}$ the 
internal photon loss rate, also decreases from $2.4\times 10^5$ to $7.7\times 10^4$.

\subsection{Integration of molecular spin qubits}
\label{ssec:synthesis}

The samples of molecular spins used in this work consist of the organic free radical 
PTMr \cite{Armet1987,Dai2018}, whose molecular structure is 
shown in Fig. \ref{fig:setup}a, diluted into a polystyrene (PS) 
polymeric matrix, with relative PTMr weight concentrations (w/w) 
ranging from $0.01$ to $5$ \%. The radical was synthesized 
using methods reported elsewhere \cite{Schafter2023}. 
The PTMr and polystyrene solutions were prepared separately in chlorobenzene, 
stirring for $1$ hour at $80$ºC before mixing to ensure a proper solubility. 
The radical samples were prepared, deposited with a micropipette and left to dry 
onto the LERs under red light, due to the photosensitivity of the radical in a liquid medium. 
The transferred volumes were $0.1 \mu$L for low-inductance LERs and $1 \mu$L for 
high-inductance LERs. After the evaporation of the solvent, the drops leave PTMr/polymer layers 
that are stable under ambient light and which cover the LER surface, as shown in 
Fig. \ref{fig:LERs}a, with a number of spins determined by the volume transferred 
and by the solution concentration.  

The morphology and topography of some of these 
deposits were studied by atomic force microscopy. 
Illustrative results are shown in 
Fig.~\ref{fig:setup}b. In all cases, 
we find that the sample thickness is smaller than 
1$0 \mu$m, which lies within the height of the 
magnetic mode generated by the LERs. 
This implies that nearly all molecules covering the 
inductor effectively interact with the photon 
magnetic field. Using these data and the ratio of 
PTMr vs polymer matrix, we estimate the number $N$ of 
spins that are coupled to each LER. The number of 
spins transferred onto the 
chip ranges from $\sim 10^{12}$ to $\sim 10^{14}$. In the case of 
high-inductance LERs, nearly all 
of them interact with the inductor. For low-inductance LERs, only 
about $1$\% are sufficiently close to the inductor wire, thus $N$ 
then ranges between $\sim 10^{10}$ and $\sim 10^{12}$ spins. Varying 
PTMr concentration allows studying how the spin-photon 
coupling depends on inductor design for comparable values of $N$. 
Regions located close to a constriction, such as 
that shown in Fig. \ref{fig:frec before after fib and fields}, behave differently. 
The field simulations (Fig. \ref{fig:frec before after fib and fields}d) show that 
only those molecules that are not much farther 
away from the constriction than its width $w$   
feel a significantly enhanced $b_{\rm mw}$. Considering a homogeneous 
distribution of PTMr over the device, as optical,
SEM and AFM images seem to indicate (Figs. \ref{fig:LERs}a, \ref{fig:frec before after fib and fields}b and \ref{fig:setup}b), gives the effective $N \sim 10^{3}$ spins 
in this case.

\subsection{Set-up for microwave transmission experiments}
\label{ssec:continuous set-up}

Figure \ref{fig:setup}c shows a scheme of the experimental setup used for most of 
the microwave transmission 
experiments described below. The superconducting devices were thermally anchored to 
the mixing chamber of a 
dilution refrigerator with a $10$~mK base temperature and 
placed inside the bore of a $1$ T cryo-free superconducting magnet whose magnetic 
field was parallel to the chip long axis. Some experiments were also 
performed using an adiabatic demagnetization refrigerator ($T \gtrsim 50$ mK) 
equipped with a home-made superconducting vector magnet. The chips were connected to 
two or four cryogenic coaxial lines that, depending on the particular experiment, 
incorporated different levels of attenuation and of cryogenic amplification.

%%%%%%%%%%%%% fig 8
%
\begin{figure}[t!]
\begin{centering}
\includegraphics[width=0.9\columnwidth, angle=-0]{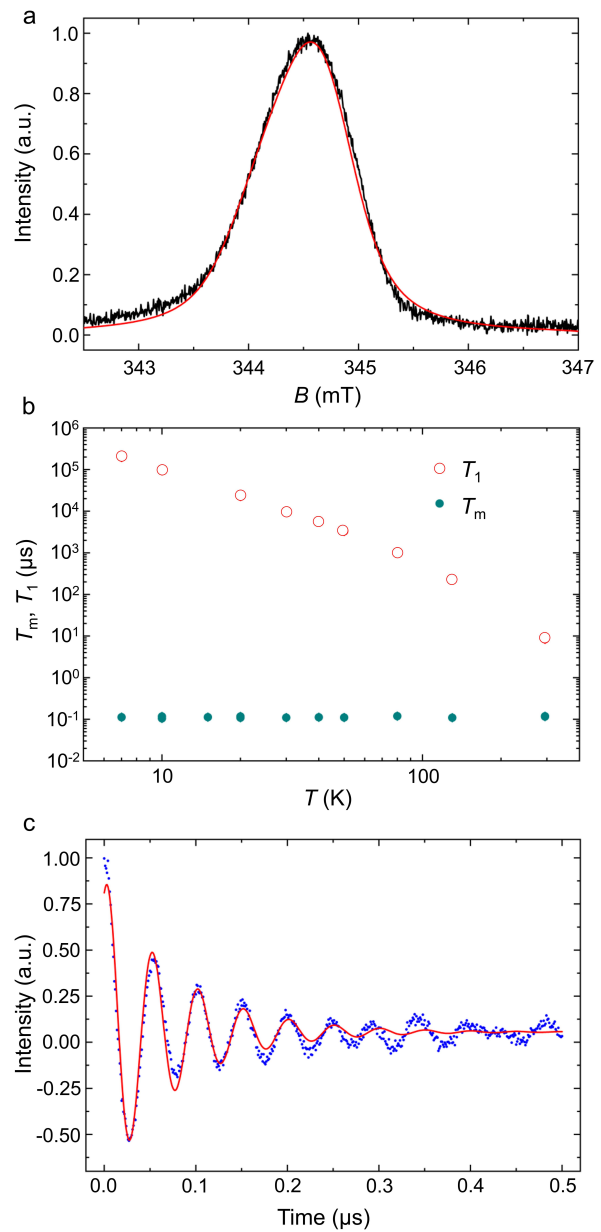}
\caption{(a) Echo-induced EPR absorption spectrum measured at 
$T=6.5$~K on a $0.5$ \% w/w solid deposit of PTMr free 
radicals embedded in a PS polymer matrix. The line is an EasySpin \cite{Stoll2006} fit that 
accounts for the line width introducing a slight anisotropy in the PTMr $g$-factor ($g_{\perp} \simeq 2.0023$ and 
$g_{z} \simeq 2.0073$). However, the inhomogeneous broadening might also have contributions from 
intramolecular hyperfine couplings \cite{Schafter2023}. (b) Temperature 
dependence of the spin phase $T_{\rm m}$ and spin relaxation $T_{1}$ times of PTMr 
derived from time-resolved EPR experiments. (c) Rabi oscillations measured at $7$~K 
with a $+3$~dB power attenuation. The solid line is a least-square fit 
with an exponentially damped oscillation $I=I(0)\exp(-t/\tau_{\rm R}) \cos\left( {\omega_{\rm R}t+\phi}\right)$. 
It gives a Rabi frequency $\Omega_{\rm R}/2 \pi \simeq 20$ MHz and a decay time $\tau_{\rm R} \simeq 80$ ns.}
\label{fig:EPR}
\end{centering}
\end{figure}

In Continuous Wave (CW) experiments, the microwave 
transmission of the device was probed, for driving frequencies $\omega/2 \pi$ ranging from  
$1$ to $8$ GHz and microwave output powers between $-45$ and $0$~dBm, 
with a Vector Network Analyzer (VNA).The input microwave signal (`readout IN' in Fig. \ref{fig:setup}c) 
was then further attenuated by $-50$~dB before reaching the device, then the transmitted signal 
(`readout OUT') was cryogenically amplified by $+30$~dB 
using a cryogenic amplifier before reaching the VNA 
detection port.

Pump-probe experiments were performed by sending sequences of excitation (typically 
tuned close to the average spin resonance frequency) and readout 
(scanning frequencies close to $\omega_{\rm r}$) pulses to the chip. In order to 
work with stronger excitation pulses, chips with separate excitation and readout 
transmission lines were used (see 
Fig. \ref{fig:setup}d and e). In this way, different net attenuations 
can be chosen for the 
pump and probe channels. As with CW measurements, the 
readout line was amplified at $T = 4$~K with a $+30$ 
dB cryogenic amplifier. The generation stage consists 
of an Arbitrary Waveform Generator (AWG) plus all the 
necessary microwave elements, such as attenuators, 
amplifiers, switches, splitters/combiners, to 
have the desired power for each of the excitation and 
readout pulses. The detection 
stage consists of a mixer, whose reference local 
oscillator signal was delivered by 
the AWG, and of a fast digital oscilloscope that 
detects  the in-phase ($I$) and quadrature ($Q$) 
components of the readout pulses.

\section{Experimental results and discussion}
\label{sec:expresults}

\subsection{PTMr free radicals as spin qubits: spin states and dynamics}
\label{ssec:EPR}

PTMr is a very stable free radical 
molecule. Its magnetic behavior is associated to an unpaired electron in its central 
carbon atom. It therefore 
provides a close realization of a model electronic spin qubit with $S = 1/2$ and a 
nearly isotropic $g$-factor $g \simeq 2.00$ \cite{Schafter2023,Dai2018,Armet1987}. When sufficiently 
isolated (i.e. in diluted solutions within nuclear spin 
free CS$_{2}$) PTMr molecules can show 
remarkably long spin coherence times, up to $150$~$\mu$s below $100$ K and close to 
$1$~$\mu$s even at room temperature \cite{Dai2018,Schafter2023}, thus being a 
promising candidate for quantum technologies.

In experiments with the superconducting circuits, we work with solid films of 
relatively concentrated PTMr samples. These 
samples have been characterized by X-band  
CW ($9.46$ GHz) and pulse ($\sim 9.66$ GHz) Electron Paramagnetic Resonance 
(EPR) experiments. For this, $400 \mu$L of a $0.5$ \% w/w PTMr/PS solution prepared as described in 
section \ref{ssec:synthesis} above were directly deposited 
into the EPR quartz tubes. Illustrative EPR results are 
shown in Fig. \ref{fig:EPR}.

The echo-induced EPR spectrum (Fig. \ref{fig:EPR}a) 
confirms that, regardless of their random orientation, PTMr free radicals are 
characterized by a well 
defined resonance frequency 
$\Omega_S / 2 \pi B = g \mu_{\mathrm{B}} / h \simeq 28$~GHz/T associated with 
the transition between spin projections $\ket{-1/2}$ and $\ket{+1/2}$ split by 
the magnetic field $B$. The inhomogeneous line 
width $\gamma/2 \pi \simeq 14$ MHz can be associated with the weak anisotropy 
in $g$ and with the hyperfine coupling to 
the $15$ Cl$^{-}$ ions ($I=3/2$) bound to the aromatic 
rings (see Fig. \ref{fig:setup}a).

%
%%%%%%%%%%%%% fig 9
%
\begin{figure*}[t!]
\begin{centering}
\includegraphics[width=1.98\columnwidth, angle=-0]{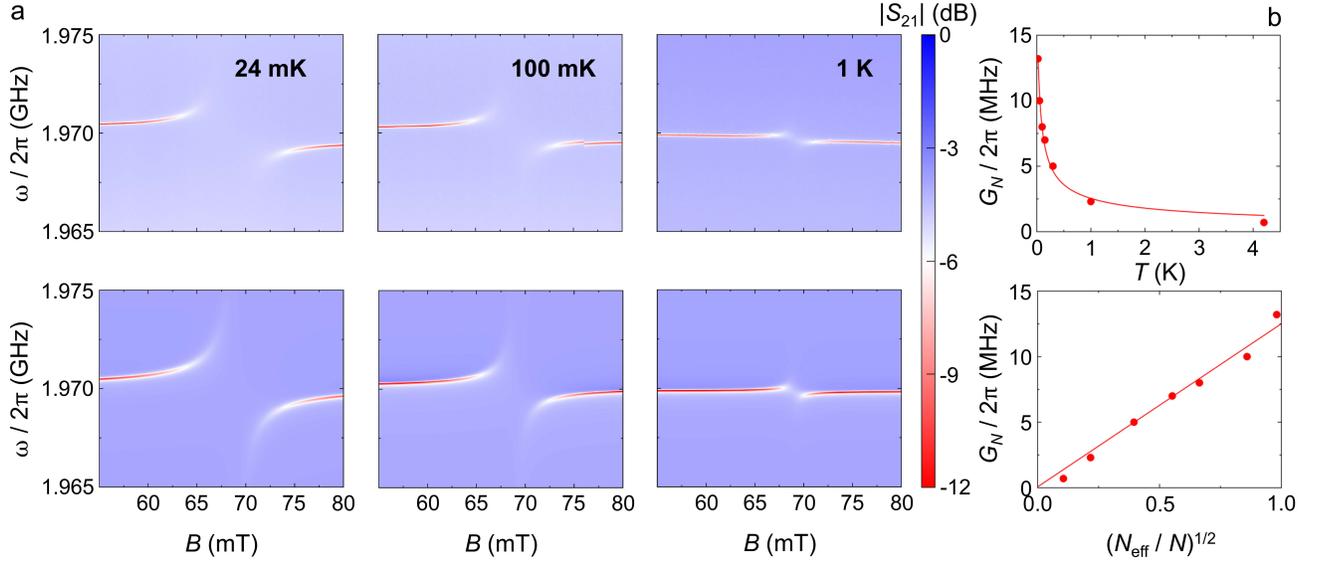}
\caption{(a) $2D$ color plots of the microwave transmission 
measured near the resonance of a $1.970$ GHz Nb LER at 
three different temperatures (top), together with 
the fits (bottom) from which the collective spin-photon 
coupling $G_{N}$ to $N \sim 5 \times 10^{12}$ PTMr free 
radical molecules was determined. (b) Temperature 
dependence of $G_{N}$. The bottom panel shows the same data 
plotted as a function of the population 
difference $N_{\rm eff}$ between the ground and excited 
spin states. Dots: experimental data; solid lines: least 
square fits based on Eq. (\ref{eq:GN_Vs_T}).}
\label{fig:2DvsT}
\end{centering}
\end{figure*}

The phase coherence $T_{\rm m}$ and spin lattice relaxation $T_{1}$ times 
have been measured by using conventional two- and three-pulse 
sequences, respectively \cite{Schafter2023}, at fixed $B = 344.5$ mT and as a 
function of temperature. They are shown in 
Fig. \ref{fig:EPR}b. The effect of concentration introduces dipole-dipole 
interactions between different free radicals, which in this 
case limit $T_{\rm m}$ to quite modest values below $200$ ns. They 
are still long enough to enable the observation of coherent Rabi 
oscillations (Fig. \ref{fig:EPR}c). The spin relaxation 
time increases with decreasing $T$, approaching $1$ s near liquid 
Helium temperatures.

\subsection{Optimizing the coupling of LERs to molecular spin qubits}
\label{ssec:coupling}

\subsubsection{Microwave transmission experiments: strong collective spin-photon coupling and optimization of its visibility}
\label{sssec:transmission}

Figure \ref{fig:2DvsT} shows $2D$ color plots of the microwave transmission measured 
near the resonance of a high-inductance 
$1.97$ GHz LER  (see Fig. \ref{fig:LERs}b) covered by 
$\sim 5 \times 10^{12}$ PTMr molecules embedded in PS. The data show clear 
signatures of the coupling of the photon and spin excitations when they are brought 
into resonance by the external magnetic field. Measurements were 
performed at different temperatures ranging from $24$ mK up to 
$1$ K. The coupling visibility becomes enhanced by cooling, as 
expected since the population difference $N_{\rm eff}$ between the 
ground and excited spin levels also increases. The spin-photon 
coupling constant $G_{N}$ was determined by fitting these data 
using input-output theory. The results, shown in 
Fig. \ref{fig:2DvsT}b, illustrate the collective spin-photon 
coupling enhancement that can be described by the well-known equation \cite{Tavis1968}

\begin{equation} \label{eq:GN_Vs_T}
	G_N = G_1\sqrt{N_{\mathrm{eff}}} = G_1\sqrt{N\tanh{\left(\frac{\hbar \Omega_S}{2k_{\mathrm{B}}T}\right)}}	
\end{equation}

\noindent where $G_{1}$ is the average coupling to individual spins. Decreasing 
temperature then allows monitoring how $G_{N}$ depends on spin polarization, or 
equivalently on $N_{\rm eff}$, as the 
bottom panel in Fig. \ref{fig:2DvsT}b shows. In addition, the results show that 
PTMr free radical spins remain paramagnetic down to very low temperatures and 
that intermolecular interactions are very weak, in spite of the relatively high 
spin concentrations used in these experiments. Likely, 
the polymer matrix efficiently prevents the formation of large molecular aggregates. 
The same fits allow estimating the inhomogeneous spin line width $\gamma$, which 
is of the order of $10-15$ MHz at all temperatures, thus 
in good agreement with the values derived previously from EPR (Fig. \ref{fig:EPR}a).

The maximum $G_{N}$ value measured at $24$ mK, 
where $N_{\rm eff} \simeq N$, amounts 
to $13.2$ MHz, which therefore borders the strong coupling 
condition $G_{N} > \{\kappa, \gamma \}$. Yet, 
the transmission at resonance $B \simeq 72$ mT does not 
show any hint of the double 
resonance dip that would characterize the onset of two non-degenerate polariton excitations. 
Rather, the transmission dip visibility becomes very small, and falls below the 
experimental noise. 

%%%%%%%%%%%%% fig 10
%
\begin{figure}[h!]
\begin{centering}
\includegraphics[width=1.0\columnwidth, angle=-0]{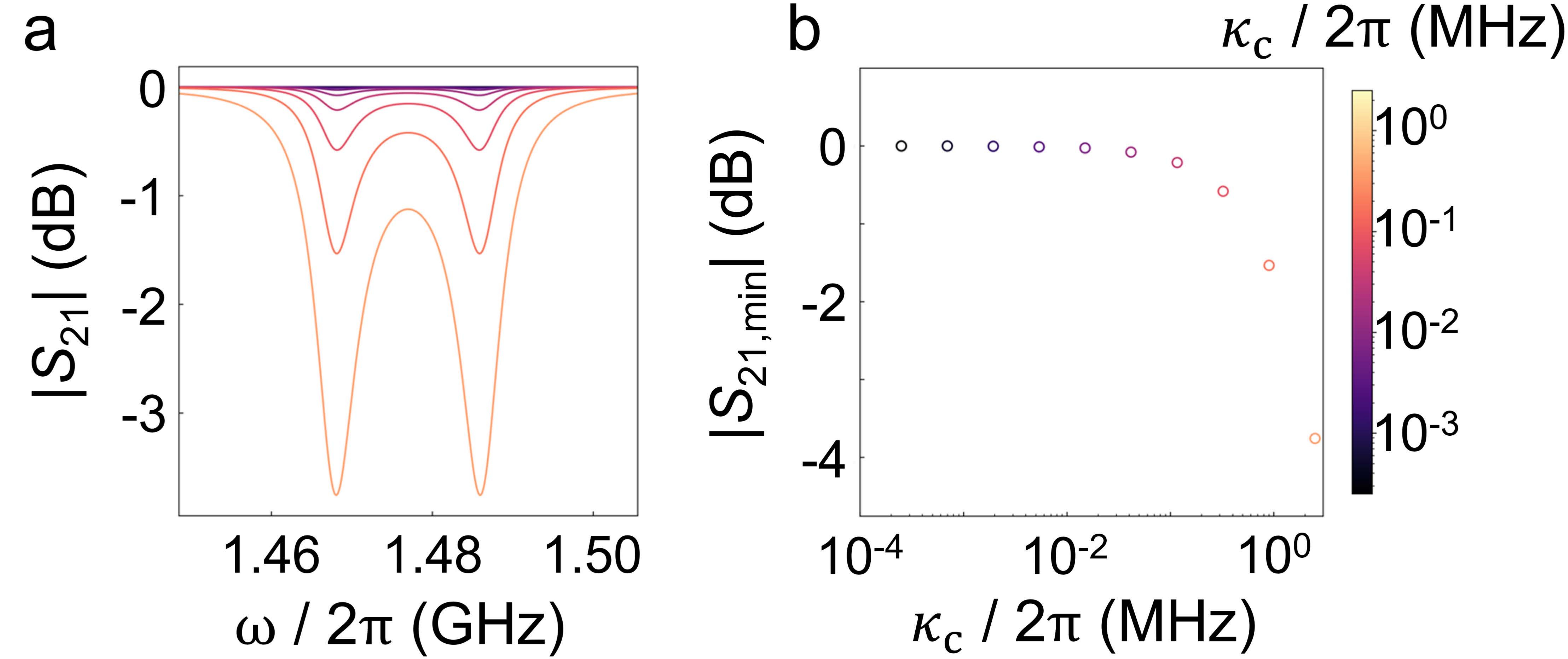}
\caption{(a) $S_{21}$ transmission parameter simulated for 
strong coupling conditions ($\kappa_{\rm i}/2 \pi = 25$~kHz, $G_{N}/2\pi = 9.5$~MHz, $\gamma/2\pi = 5$~MHz), showing 
how the visibility of the transmission dips associated 
with the excitation of the two polariton modes increases with $\kappa_{\rm c}$. (b) 
Evolution with $\kappa_{\rm c}$ of the transmission dip amplitudes shown in (a).}
\label{fig:visibility theory}
\end{centering}
\end{figure}

A simple strategy to enhance the polariton visibility in transmission is to adjust 
$\kappa_{\rm c}$ by design. Figure~\ref{fig:visibility theory} shows 
theoretically that by increasing 
$\kappa_{\rm c}$ the two transmission resonance depths can be tuned for any 
given $G_{N}$ and $\gamma$ values, thus overcoming the noise level. To validate this 
approach experimentally, we compare the coupling of the same PTMr deposit to the 
ground and first excited modes of a LER. In this case, the change 
in $\kappa_{\rm c}$ is caused by the different current 
distributions of these two modes, which determines its interaction with the readout transmission 
line. Measurements of the bare LER response performed at $B=0$ show 
that its first resonant mode has $\omega_{\rm r} / 2 \pi = 1.478$ GHz and 
$\kappa_{\rm c} / 2 \pi = 113$ KHz, whereas the second 
resonates at a higher $\omega_{\rm r} / 2 \pi = 6.265$ GHz with a much 
higher $\kappa_{\rm c} / 2 \pi = 3.175$ MHz.

%%%%%%%%%%%%% fig 11
%
\begin{figure}[h!]
\begin{centering}
\includegraphics[width=1.0\columnwidth, angle=-0]{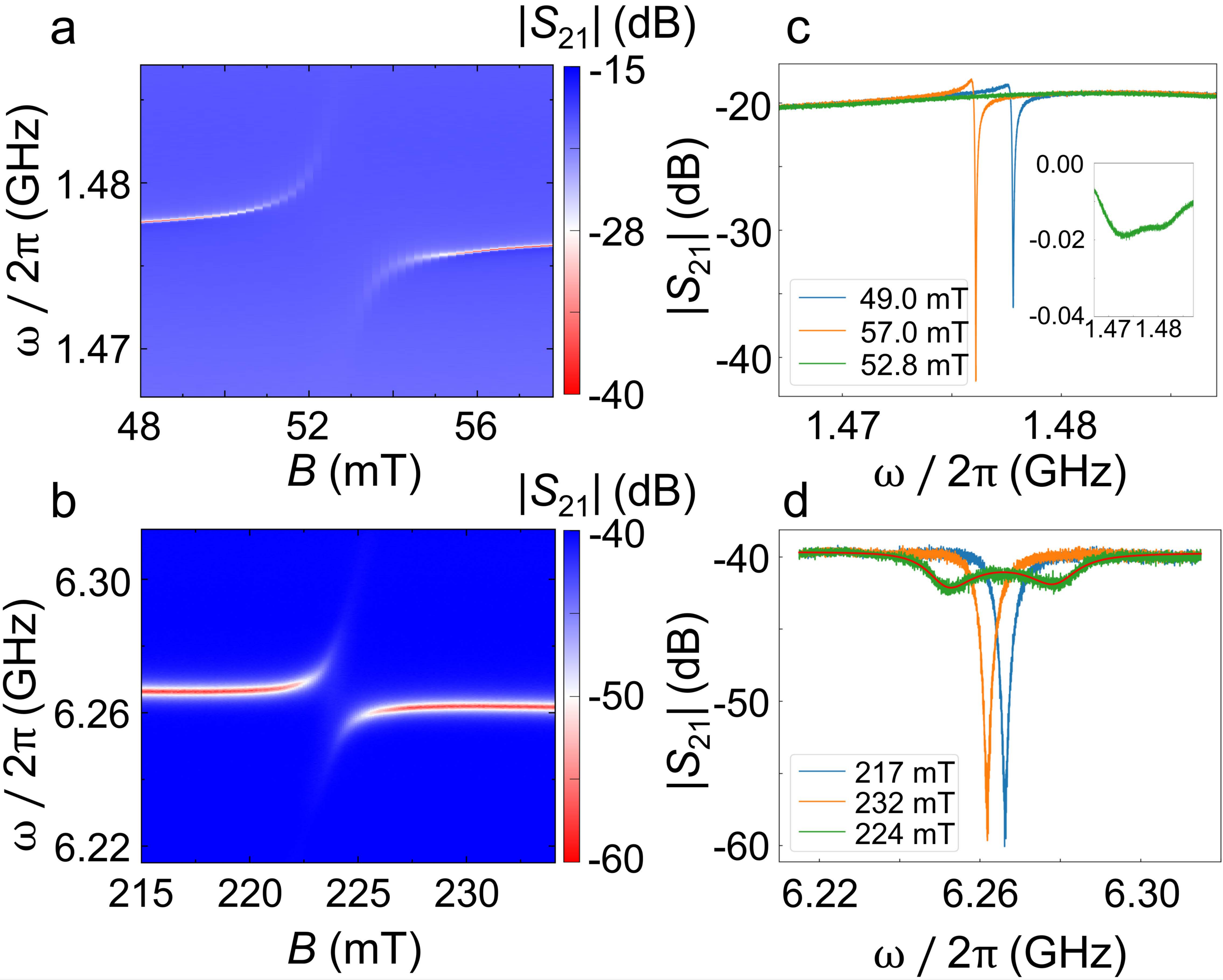}
\caption{(a) and (b) $2D$ color plots of the microwave 
transmission measured at $T=10$ mK near the ground mode 
resonance of a NbTiN $1.477$ GHz LER having $\kappa_{\rm c}/ 2 \pi = 113$ 
KHz and near the first excited mode of the same LER with 
$\omega_{\rm r}/2 \pi = 6.265$ GHz and $\kappa_{\rm c}/ 2 \pi = 3.175$ MHz, respectively. 
The LER was coupled to $N \sim 10^{12}$ PTMr molecules embedded in a PS matrix. 
(c) $S_{21}$ transmission 
parameter measured as a function of driving frequency 
$\omega$ at three different magnetic fields near the spin 
resonance ($52.8$ mT) with the ground photon mode. 
Inset: $S_{21}$ transmission measured at $52.8$~mT 
corrected from the $S_{21}$ transmission background measured at 
zero field. (d) $S_{21}$ transmission parameter measured as 
a function of driving frequency $\omega$ at three different 
magnetic fields near the spin resonance field ($224$~mT) 
with the first excited mode. The red line is the strong 
coupling fit.}
\label{fig:visibility}
\end{centering}
\end{figure}

Field dependent $2D$ transmission color plots for these 
modes, measured at $T=10$ mK, are shown in 
Figs.~\ref{fig:visibility}a and b. In both cases, clear 
signatures of the coupling to the spins are seen. For the 
ground mode, a $2D$ fit similar to those shown in Fig.~\ref{fig:2DvsT}a yields a 
collective spin-photon coupling $G_N / 2 \pi= 9.5$~MHz and a spin line width 
$\gamma / 2 \pi = 5$~MHz. Even though $G_{N} > \gamma$, the two anticrossing 
branches tend to fade away close to the spin-photon resonance magnetic field 
$\simeq 52.8$ mT. Figure \ref{fig:visibility}c shows frequency-dependent 
transmission data measured at three selected magnetic fields: below resonance 
($48$~mT), at resonance ($52.8$~mT), and above resonance ($57$~mT). None of them 
provides any clear evidence for the two peaks associated 
with the polaritonic excitations, even after carefully 
subtracting the background signal (inset in Fig. \ref{fig:visibility}c). 
The very low visibility 
also hinders a precise determination of $G_N$ and of 
$\gamma$. The situation becomes qualitatively different 
when the spins are brought to resonance, near $223.9$ mT, with the excited mode of 
the same LER. In this case, the two peaks become clearly visible, see 
Fig.~\ref{fig:visibility}d, which allows a straightforward 
fit yielding $G_N / 2 \pi= 15$ MHz and $\gamma / 2 \pi = 12.3$~MHz.

In these experiments, we are dealing with large mode volumes and correspondingly 
large ensembles of $N \sim 5\times 10^{12}$ spins. The average coupling to each 
individual spin can then be estimated from Eq. (\ref{eq:GN_Vs_T}), as shown in 
Fig. \ref{fig:2DvsT}. The obtained $G_1/2 \pi = 6 \pm 0.4$~Hz is 
somewhat larger than the coupling strengths of free radicals to conventional co-planar 
waveguide resonators, which are typically below $3-4$ Hz for comparable 
frequencies \cite{Gimeno2020}. Yet it is 
still much lower than the inhomogeneous spin line width 
$\gamma/2 \pi \sim 10-15$ MHz, and than the spin 
decoherence rates of PTMr embedded solid films 
($1/2\pi T_2 \sim 1.2$ MHz, Fig. \ref{fig:EPR}) and of 
optimally isolated PTMr in solution ($1/2\pi T_{2} \sim 1$ 
kHz, see Ref. \onlinecite{Schafter2023}). 
Reaching the strong coupling regime with a single 
spin is therefore very challenging 
unless the microwave magnetic field is significantly 
enhanced. A possible approach to 
attain this goal is introduced in the following 
section.

\subsubsection{Enhancing the single spin-photon coupling in low-inductance LERs}
\label{sssec:low-L-LERs}

As it has been shown in section \ref{ssec:LERs}, low-$L$ LERs are the 
starting point for enhancing the coupling to each 
individual spin. By reducing the inductor to just a 
single straight wire, the inductance 
can be reduced by a factor $10$. We have studied the 
chip shown in Fig. \ref{fig:LERs}a, which contains 
$10$ such LERs. In half of them, a $w=50$ nm wide 
$500$ nm long nano-constriction was fabricated at the 
center of the inductor wire, in order to further 
enhance $b_{\rm mw}$ in its close neighborhood 
(see Fig.~\ref{fig:frec before after fib and fields}d). We 
have studied their coupling to PTMr samples with 
similar geometries but different concentrations, in 
order to vary $N$.

%
%%%%%%%%%%%%% fig 12
%
\begin{figure}[t]
\begin{centering}
\includegraphics[width=0.9\columnwidth, angle=-0]{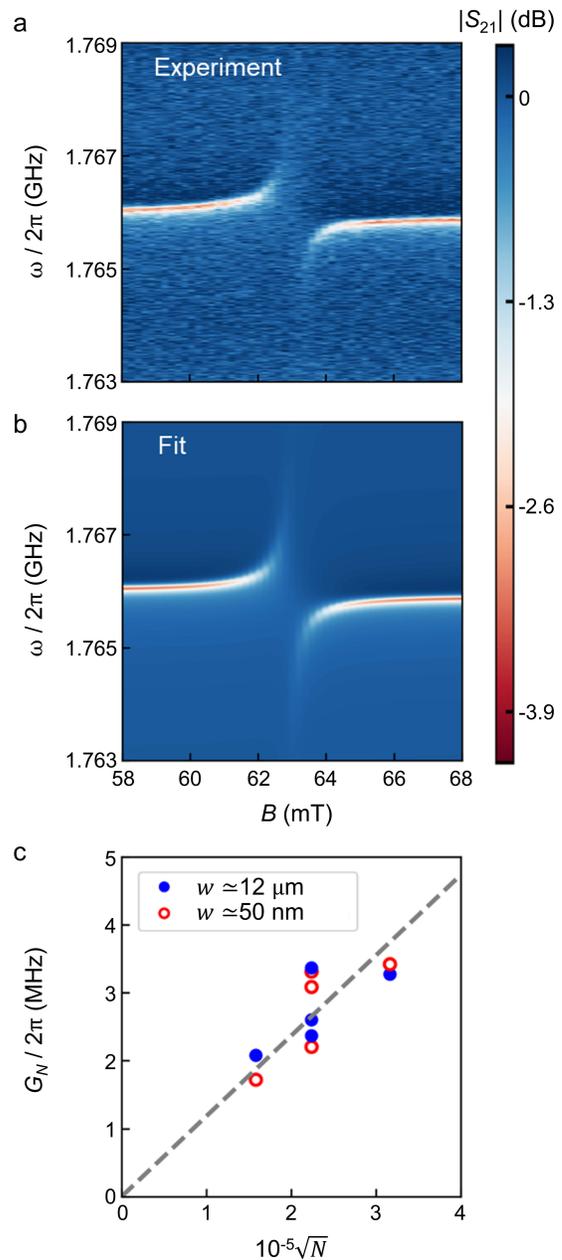}
\caption{(a) Microwave transmission measured at $T=10$ mK 
near the resonance of a $\omega_{\rm r}/2 \pi = 1.766$ GHz 
low-inductance Nb LER. (b) Theoretical fit of the 
transmission based on input-output theory. (c) Dependence 
of the collective spin-photon coupling $G_{N}$ 
measured on low-inductance LERs at $T=10$ mK as a function 
of the number of spins $N$ deposited onto the inductor.}
\label{fig:lowLresults}
\end{centering}
\end{figure}

Figure~\ref{fig:lowLresults} shows microwave transmission 
data measured at $T= 10$ mK near the resonance of one of 
the LERs with the PTMr spins and the $2D$ fit that 
enables estimating $G_{N}$. The coupling strengths range 
between $2$ and $4$ MHz, 
which are smaller than those found for the high-inductance 
LERs on account of the 
smaller number of molecules that are effectively coupled to the single wire 
inductor. As shown by Fig.~\ref{fig:lowLresults}c, $G_{N}$ mainly 
correlates with $N$. From this plot, we estimate the average coupling to 
individual spins $G_{1} \sim 12$ Hz, which is a 
factor $2$ higher than the values found for high-$L$ 
LERs. However, for any given $N$ and within the 
experimental uncertainties we do not find significant 
differences associated with the nanoconstrictions. 
This can be explained by the homogeneous filling of 
the LER electromagnetic mode by $10^{10}-10^{11}$ 
spins and from the minute 
fraction of molecules located sufficiently close to 
the nanoconstriction of each LER (at most $10^3$), 
which largely compensates for their enhanced $G_{1}$. 
While the average single spin coupling remains modest, this experiment 
establishes the baseline for evaluating local-field 
enhancements, analyzed in next 
section \ref{ssec:readout} on 
the basis of dynamical measurements, and 
confirms that the inductor design provides a way to 
improve $G_{1}$.

\subsection{Pump-probe dispersive readout experiments}
\label{ssec:readout}

\subsubsection{Detection of the thermal spin polarization}
\label{sssec:thermal-shift}
In this and the following sections we consider experiments performed with sequences 
of microwave pulses, by which we explore the ability to use the circuit to modify 
the spin states and then readout the results. In most of these experiments, we set 
the spin-LER system in the dispersive regime \cite{Blais2004}, where the relevant 
frequencies of each sub-system, $\Omega_{S}$ and $\omega_{\rm r}$ are sufficiently 
detuned with respect to each other. This condition is met for 
$\vert \Delta \vert \equiv \vert \Omega_{S} - \omega_{\rm r} \vert \gg G_{N}$.

Even though no real exchange of excitations between both systems occurs in this 
regime, their frequencies are still affected by their mutual coupling, meaning that 
$\omega_{\rm r}$ becomes sensitive to the spin 
polarization. When the latter is nonzero, $\omega_{\rm r}$ is pushed away from 
the bare resonance frequency $\omega_{\rm r,0}$ according 
to the following expression:
\begin{equation} \label{eq:ensemble_dispersive_shift}
	\omega_{\rm r} = 
	\omega_{\rm r,0} + \sum_{j=1}^N \chi_j
	\langle \sigma_{z,\,j} \rangle
	\textrm{,}
\end{equation}
\noindent where 

\begin{equation} \label{eq:single_dispersive_shift}
	\chi_{j} = \frac{G_{j}^{2}\Delta_{j}}{\Delta_j^2 + (1/T_2)^2}
	\textrm{,}
\end{equation}
is the maximum shift generated by the $j$-th spin in the 
ensemble, associated with a spin polarization 
$\langle \sigma_{z,j} \rangle = -1$, and $\Delta_{j} \equiv \Omega_{S,j}-\omega_{\rm{r},0}$ 
gives the detuning of its resonance frequency $\Omega_{S,j}$ with respect to that of the LER. 
In the experiments, the spin system is initially in its thermal 
equilibrium state, characterized at any temperature $T$ and 
magnetic field $B$ by 
$\langle \sigma_{z,j} \rangle_{T} \simeq -\tanh \left(g \mu_{\rm B} B S/k_{\rm B} T \right)$. 

%%%%%%%%%%%%% fig 13%
\begin{figure}[t!]
\begin{centering}
\includegraphics[width=0.98\columnwidth, angle=-0]{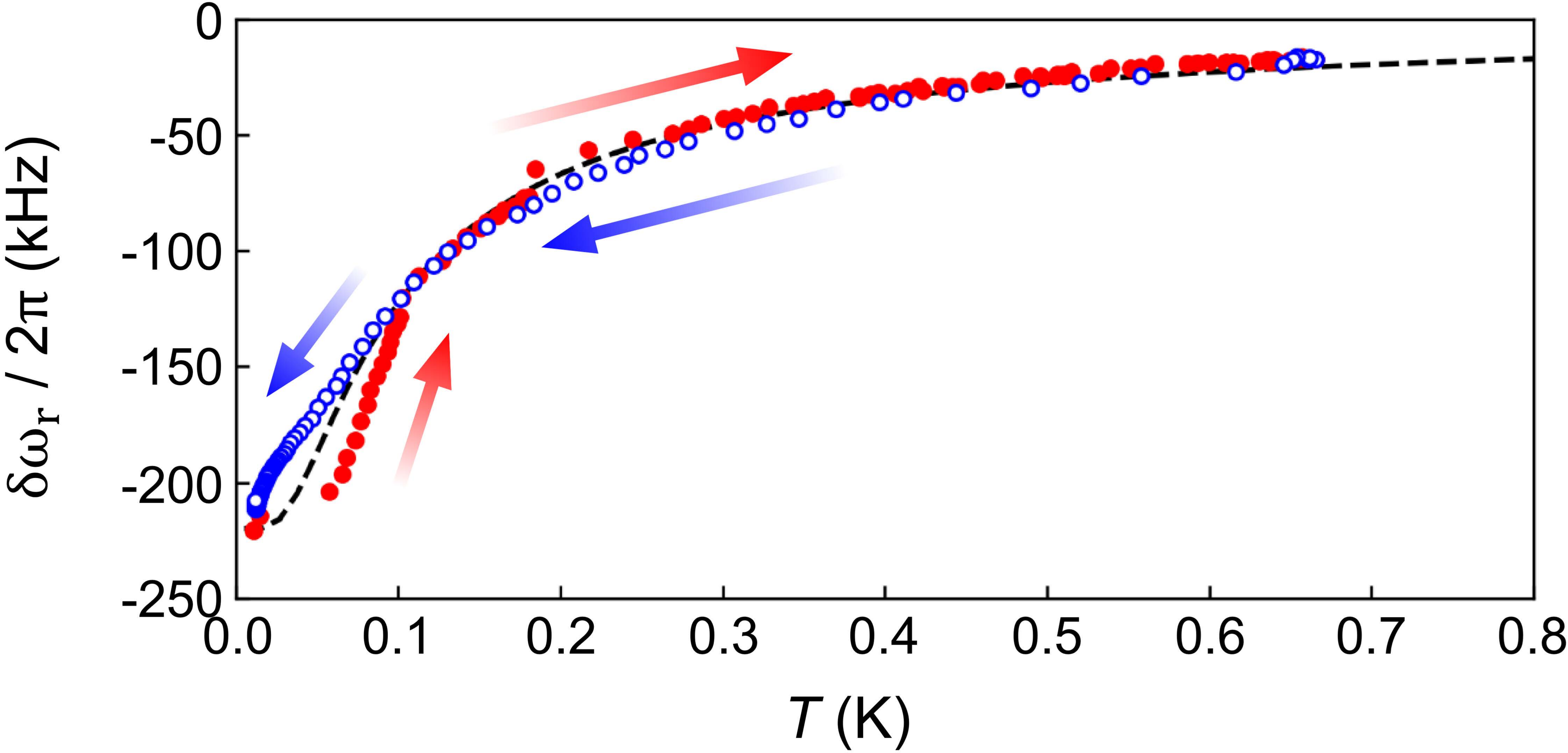}
\caption{Resonance frequency shift $\delta \omega_{\rm r} = \omega_{\rm r}-\omega_{\rm r,0}$ of a 
$2.564$ GHz low-inductance Nb LER generated by its coupling to $N \simeq 10^{12}$ PTMr 
free radical molecules, measured as a function of 
increasing (red arrows) and then decreasing 
(blue arrows) temperature under a static magnetic 
field $B = 93.7$ mT. The dots are the experimental 
data and the dashed line follows from 
Eq. (\ref{eq:ensemble_dispersive_shift}) for 
thermal spin polarizations.}
\label{fig:shiftvsT}
\end{centering}
\end{figure}

The link between $\omega_{\rm r}$ and $\langle \sigma_{z} \rangle_{T}$ can be experimentally 
tested by measuring the LER resonance as a function of temperature. Results measured at 
$B = 93.7$ mT for a $2.564$ GHz LER are shown in Fig. \ref{fig:shiftvsT}. 
Under these conditions, the average $\Omega_{S}/2 \pi \simeq 2.604$ GHz, thus $\Omega_{S} > \omega_{\rm r,0}$ and the 
spins exert an increasingly negative shift $\delta \omega_{\rm r} \equiv \omega_{\rm r} - \omega_{\rm{r},0}$ on 
the LER frequency as they become progressively polarized on cooling. Its temperature dependence resembles 
Curie's law. The deviation from the equilibrium predictions observed below $40$ mK 
is associated with a slowing down of the spin-lattice 
relaxation (see below). Even so, the spin polarization 
reaches values close to $1$ at the base temperature $T \simeq 11$ mK.

\subsubsection{Dispersive measurement of the spin excitation spectrum}
\label{sssec:dispersive_excitations}
\begin{figure}[t!]
\begin{centering}
\includegraphics[width=0.98\columnwidth, angle=-0]{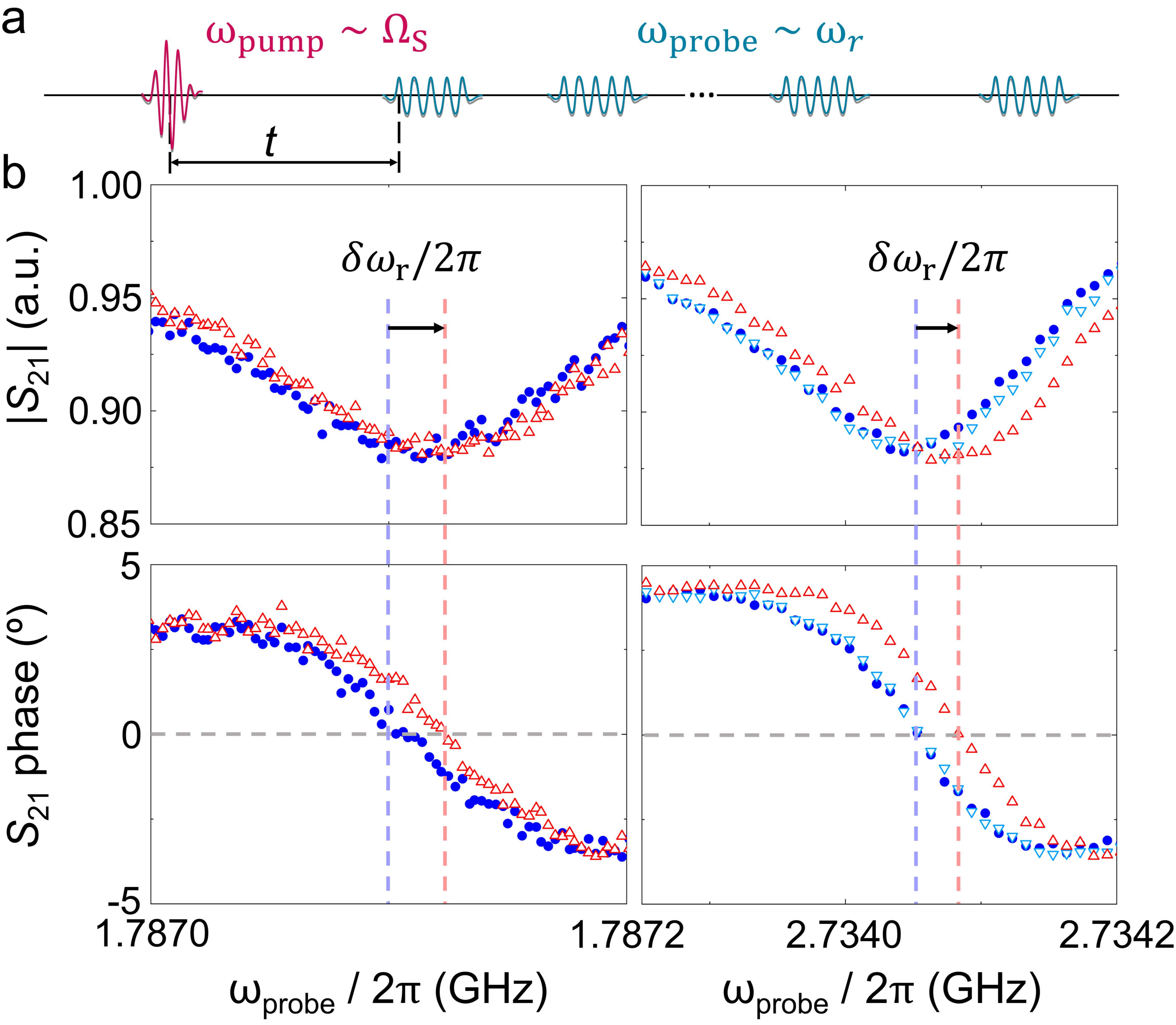}
\caption{(a) Scheme of the pulse sequence applied to 
the superconducting circuit in order to measure the dispersive shift 
$\delta \omega_{\rm r}$ associated 
with changes in the spin states. The excitation of 
the spin ensemble with a pump pulse of frequency 
$\omega_{\rm pump} \sim \Omega_{S}$ is 
followed, after a time $t$, by a train of $\sim 100$ readout 
pulses and signal acquisitions at different frequencies $\omega_{\rm probe}$ near 
$\omega_{\rm r}$, which detect changes in the LER 
response. (b) Response amplitude (top) and phase (bottom)
measured at $T=10$~mK near $\omega_{\rm r}$ of two low-$L$ LERs on chips with a single 
excitation/readout line (left) and with two 
independent lines (right). Dark blue dots: resonance 
measured before exciting the spin transition; open red 
triangles, resonance measured after exciting the spins 
with a pulse of frequency $\omega_{\rm pump} = \omega_{\rm r} + \Delta \simeq \Omega_{S}$; 
light-blue inverted triangles, resonance measured after 
sending a pulse with a frequency 
$\omega_{\rm pump} = \omega_{\rm r} - \Delta$, thus equally 
detuned from the resonator but not driving the spins.}
\label{fig:pump-probe}
\end{centering}
\end{figure}

The coupling to the circuit can also induce and detect 
deviations of the spins from thermal equilibrium. The sequence of 
microwave pulses typically includes spin excitation 
pulses, with frequencies close to the average  
$\Omega_{\rm S}$, followed by another sequence of readout 
pulses, with frequencies close to $\omega_{\rm r}$, that 
detect any shift $\delta \omega_{\rm r}$ in the LER resonance. Figure \ref{fig:pump-probe}a shows 
a scheme of the pulse sequences and examples of LER resonances measured before 
and after pumping on the spins, in this case for an average $\Delta > 0$. 
The long spin-lattice relaxation times allow scanning the whole resonance 
before the decay of spin excitations. These excitations 
modify $\langle \sigma_{z,j} \rangle$ and, 
following Eq. (\ref{eq:ensemble_dispersive_shift}), shift 
the LER frequency from its reference value, corresponding 
to the equilibrium spin state, to a higher value. Yet, $\delta \omega_{\rm r}$ is barely 
visible above the noise level in the experiments performed with a single 
excitation/readout transmission line (Fig. \ref{fig:pump-probe}b, left).

%%%%%%%%%%%%% fig 14

The frequency shift can be enhanced by applying 
stronger spin pumping pulses on chips which include 
two independent transmission lines, like the one in 
Fig.~\ref{fig:setup}. The 
excitation line couples high-power ($\sim 0$ dBm) pulses 
directly to the spins and to the LER inductor, while the 
readout line couples low-power ($-50$ dBm) pulses 
only to the LER capacitor. The risk of a power leak from 
the excitation line into the readout line through the LER 
is avoided by working in the dispersive 
regime, as the LER acts as a band-pass filter. With this 
architecture, larger values of $\delta \omega_{\rm r}$ are 
attained for the same excitation pulse durations, as seen 
on the right hand panel of Fig. \ref{fig:pump-probe}b.

%%%%%%%%%%%%% fig 15
\begin{figure}[t!]
\begin{centering}
\includegraphics[width=0.95\columnwidth, angle=-0]{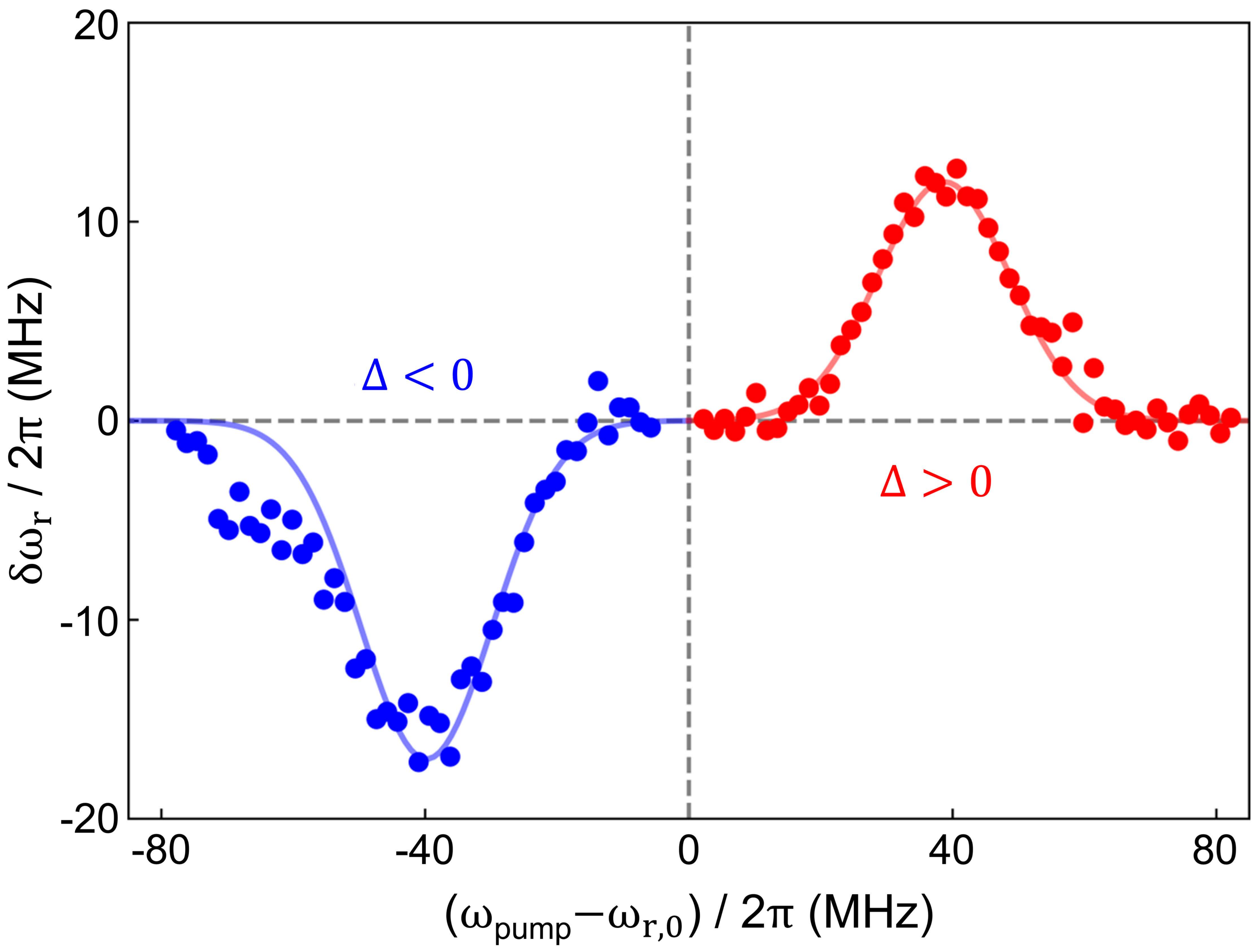}
\caption{Spin resonance spectrum of PTMr organic free 
radicals measured, at $T = 10$~mK, from the 
frequency shift $\delta \omega_{\rm r}$ of a low-$L$ Nb 
LER coupled to $N \simeq 5.5 \times 10^{11}$ molecules. The shift is generated 
by spin excitations induced by the application of 
$50 \mu$s long pump 
pulses with varying frequencies $\omega_{\rm pump}$, close 
to the average spin resonance frequency $\Omega_{S}$. 
Solid blue and red dots correspond 
to experiments performed at $B=62.3$ mT and $B=65$ mT, respectively. 
They give rise to a negative and positive detuning $\Delta / 2 \pi \simeq  \mp 40$ MHz 
of the average $\Omega_{S}$ with respect to the bare LER frequency 
$\omega_{\rm r,0}/2 \pi = 2.754$ GHz.}
\label{fig:spectroscopy}
\end{centering}
\end{figure}

In the single spin limit, $\delta \omega_{\rm r}$ provides a `dispersive readout' of the qubit 
state, a well established technique for superconducting 
qubits \cite{Schuster2005, Wallraff2005} that has nowadays been extended to other 
qubit realizations coupled to circuit quantum electrodynamics 
platforms \cite{Scarlino2019}. Here, we are 
instead dealing with an ensemble of spins with different 
characteristic frequencies. Each pump pulse, with a finite duration 
$\sim 50 \mu$s,  
excites only the part of the distribution that lies within 
its bandwidth $\sim 1/(50 \, \mu\rm{s}) = 20$ kHz 
centered at $\omega_{\rm{pump}}$. Therefore, by 
sweeping $\omega_{\rm pump}$ a direct picture of the spin 
frequency distribution can be obtained. Dispersive 
spectroscopy has been applied to characterize molecular 
spins by means of coplanar resonators \cite{Bonizzoni2021}. The present platform 
allows scanning $\omega_{\rm pump}$ over a wider range, as these pulses 
are generated by an open transmission line.

Figure \ref{fig:spectroscopy} shows the change in $\delta\omega_{\rm r}$ measured as 
$\omega_{\rm pump}$ is swept across the average spin 
resonance frequency $\Omega_{S}$. Data measured for $B=65$mT and for $B=62.3$ mT, 
which generate opposite average detunings 
$\Delta / 2\pi \simeq \pm 40$ MHz, are shown. The sign of 
$\delta\omega_{\rm r}$ follows the sign of $\Delta$, as 
expected from Eq. (\ref{eq:ensemble_dispersive_shift}). Both spectra show a Gaussian 
line shape with $\sigma / 2 \pi = 10.01$ MHz. The half-width at half maximum 
(HWHM) $\sqrt{2 \ln{(2)}} \sigma / 2 \pi = 11.79$ MHz 
is consistent with the spin broadening $\gamma$ derived from the CW 
measurements discussed above.

\subsubsection{Relaxation of `dressed' spin states: Purcell effect and estimation of the maximum 
single spin-photon coupling}
\label{sssec:relaxation}

%%%%%%%%%%%%% fig 16
%
\begin{figure}[t!]
\begin{centering}
\includegraphics[width=0.9\columnwidth, angle=-0]{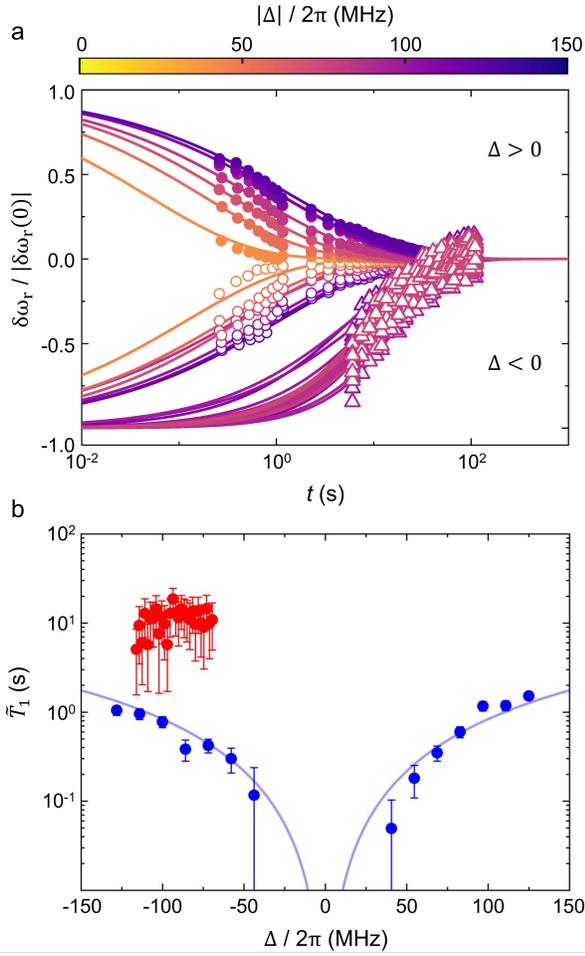}
\caption{(a) Shift in frequency $\delta \omega_{\rm r}$ 
of a $1.970$ GHz high-$L$ LER, at $T=45$ mK (open 
triangles), and of a $1.787$ GHz low-$L$ LER, at $T=10$ 
mK (circles), measured at a time $t$ after applying a pump pulse 
whose frequency is detuned an amount $\Delta$ from the bare resonator frequency 
$\omega_{\rm r,0}$. The data have been normalized by 
$\vert \delta \omega_{\rm{r}}(t \rightarrow 0) \vert$. 
Notice that $\delta \omega_{\rm r}$ changes sign with $\Delta$, as 
expected (see Fig. \ref{fig:spectroscopy}). 
The solid lines are least-square stretched exponential 
fits (Eq. (\ref{eq: 06_StretchedExponential})). 
(b) Dependence of the spin relaxation time 
$\tilde T_{1}$ extracted from these fits as a function of 
$\Delta$. Red dots, high-$L$ LER; blue dots, 
low-$L$ LER. The solid line is a least-square fit based on 
Eq. (\ref{eq:ybridSpinRelaxationRate}), which includes the 
photon-induced Purcell effect.}
\label{fig:shift-relaxation-1}
\end{centering}
\end{figure}

After the spins are excited by a pump pulse, the frequency shift 
$\delta \omega_{\rm r}$ that is generated immediately starts 
to decay back to zero. The top panel 
of Fig. \ref{fig:shift-relaxation-1} shows that 
$\delta \omega_{\rm r}(t)/\vert \delta \omega_{\rm r} (0)\vert $ 
decays with the time interval $t$ separating pump and 
readout pulses. These experiments were performed at $T \simeq 10-40$ mK 
and for average detuning $\vert \Delta \vert \gtrsim 44$ MHz. 

Relaxation was measured both with a high-$L$ and with 
a low-$L$ LERs. The experiments evidence a remarkably 
slow relaxation, which dies off only for times of the 
order of $100$ s. The relaxation is 
non exponential. Since PTMr molecules are very weakly 
interacting, this likely reflects the 
existence of a distribution in spin relaxation times in 
the ensemble. The distribution of exponential decays can 
be fitted with a stretched exponential (see the solid 
lines in the top panel of Fig. \ref{fig:shift-relaxation-1}).

\begin{equation} \label{eq: 06_StretchedExponential}
	\begin{split}
		\delta \omega_{\rm r} (t) \equiv \delta \omega_{\rm r} (0) e^{-(t/\tilde{T_1})^x}
		\textrm{,}
	\end{split}
\end{equation}

\noindent where $x \leq 1$ is the stretch parameter and 
$\tilde{T_{1}}$ is an average 
spin relaxation time. The fits also reveal that 
$\tilde{T_{1}}$ is significantly shorter, and the 
distribution broader, for the relaxation measured with a 
low-$L$ LER.

These results suggest that the spin-photon coupling, which 
depends on circuit design, modifies the spin 
thermalization speed. They can be qualitatively understood on 
the basis of the Purcell 
effect \cite{Purcell1946, Boissonneault2009, Reuther2010}, which gives each spin $j$ 
in the ensemble an additional relaxation path through its coupling $G_{1,j}$ 
to the cavity. In the dispersive regime, the spin 
hybridization with the photon modes, of the order 
of $|G_{1,j}/\Delta_{j}| \ll 1$,
introduces a net probability 
for the spin to relax via the cavity decay channels, whose 
rate $\kappa$ is typically orders of magnitude higher than the intrinsic $1/T_{1}$ 
of the bare material. Then, the `photon-dressed' 
relaxation rate $1/\tilde{T}_{1,j}$ of each spin $j$ is enhanced by an 
additional term \cite{Zueco2019}:
\begin{equation} \label{eq:ybridSpinRelaxationRate}
	\dfrac{1}{\tilde{T}_{1,j}} = \dfrac{1}{T_{1}} + \frac{4G_{1,j}^2 \kappa \omega_{\rm r,0} \Omega_{S,j}}
    {\left[ \omega_{\rm r,0}^{2} - \Omega_{S,j}^2 \right]^2 + \left(\kappa \Omega_{S,j}\right)^2}\ ,
\end{equation}
\noindent that depends on properties of each individual spin.
%%%%%%%%%%%%% fig 17
%
\begin{figure}[t!]
\begin{centering}
\includegraphics[width=0.95\columnwidth, angle=-0]{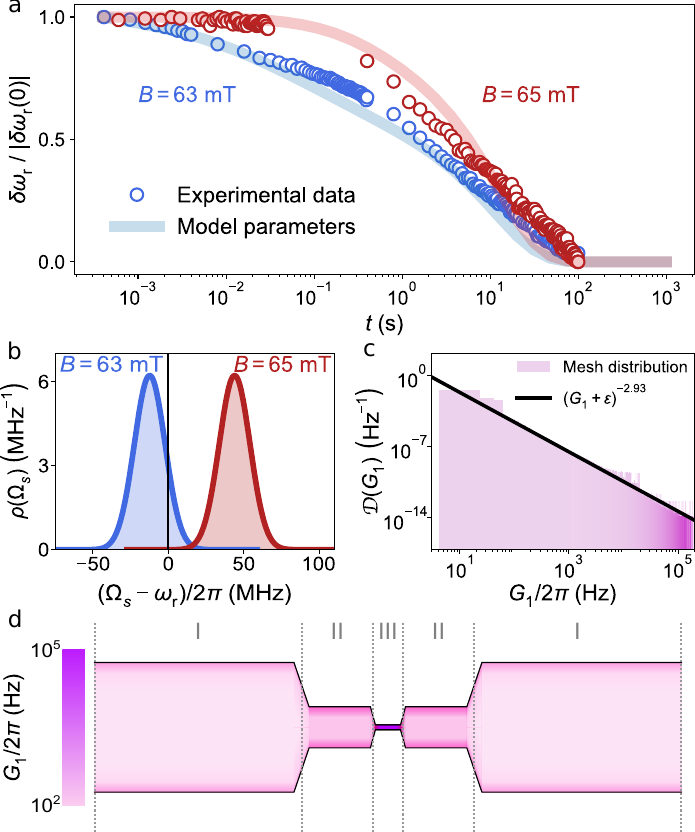}
\caption{(a) Decay of the shift in frequency of a $1.787$ 
GHz low-$L$ LER with a $50$ nm 
wide constriction in its inductor line (see Fig. \ref{fig:frec before after fib and fields}) 
following the application of a $50$ $\mu$s long pulse 
tuned to the average $\Omega_{S}$ of PTMr free-radical molecules coupled to it. 
The data were measured at $T = 10$ mK over a time window spanning six decades 
and for two different magnetic fields, which lead to different average spin-LER 
frequency detuning. The 
symbols are experimental data whereas the solid lines follow from the theoretical 
model described in the text. 
This model uses the distribution of spin frequencies shown in (b), which has been
obtained from independent 
measurements (Fig. \ref{fig:spectroscopy}). In addition, 
it introduces a power-law for 
the distribution ${\cal D}(G_{1,j})$ of single spin to 
photon couplings $G_{1,j}$, 
shown in (c), leaving the maximum $G_{1,j} \sim 100$ kHz 
as the only free parameter. The shape of this distribution follows from 
the close to cylindrical symmetry of the LER inductor. The 
coloured bars in this plot show the 
results of finite-element simulations of $G_{1,j}$  performed with the geometry 
shown in (d).}
\label{fig:shift-relaxation-2}
\end{centering}
\end{figure}

\noindent It follows from Eq.(\ref{eq:ybridSpinRelaxationRate}) that the spin-cavity 
detuning controls the effect that photons have on the spin 
wavefunctions. This effect can therefore be studied by 
measuring the dependence on $\Delta$. Relaxation curves measured, for both, high-$L$
and low$-L$ LERs and for different detunings, are shown in Fig. \ref{fig:shift-relaxation-1}a, 
whereas Fig. \ref{fig:shift-relaxation-1}b compares $\tilde{T_{1}}$ 
data obtained from stretched exponential fits. For the high-$L$ LER, 
$\tilde{T_{1}}$ depends little, within the experimental 
uncertainties, on $\Delta$. We associate this 
to the intrinsic spin-lattice relaxation of PTMr. By 
contrast, in the case of the low-$L$ LER, $\tilde{T_{1}}$ 
decreases with decreasing $\vert \Delta \vert$, signaling 
the onset of progressively faster photon induced 
relaxation.

The effectiveness of the Purcell effect can be fully 
appreciated in Fig. \ref{fig:shift-relaxation-2}a, which 
shows relaxation data obtained in an extended time region. 
Remarkably enough, there is a detectable relaxation that occurs within time 
scales of the order $10-100$ ms, and 
which speeds up as $\vert \Delta \vert$ 
decreases. Through the Purcell effect, the distribution of spin-photon couplings 
$G_{1,\,j}$ that arises from the inductor geometry (see Figs. \ref{fig:shift-relaxation-2}c 
and d) naturally accounts for the distribution of 
spin relaxation times that is observed experimentally. It also  
suggests that very few spins might be experiencing local couplings orders of 
magnitude stronger than the average derived from the continuous wave 
experiments. Notice that spin-spin interactions, which might also speed up 
relaxation, are weak in these samples and, furthermore, that they would not account 
for the dependence of $\tilde{T_{1}}$ of the spin-photon detuning and on the LER 
inductance that are observed experimentally (Fig. \ref{fig:shift-relaxation-1}). The 
relaxation rate driven by the Purcell effect then provides quantitative 
information on the couplings of individual spins to the 
LER and on their distribution across the molecular spin 
ensemble.

In order to estimate the maximum $G_{1,j}$ compatible with 
these data, we have used a simple model based on Eqs. (\ref{eq:ensemble_dispersive_shift})
and  (\ref{eq:ybridSpinRelaxationRate}) in combination 
with the available information on spin excitations and LER geometry. The decay of $\delta \omega_{\rm r}$  
has been calculated by the following expression 

\begin{equation}
    \delta\omega_r(t) = \int_{\Omega_\text{S}}\int_{G_1}\chi\langle\sigma_z\rangle(t) \rho(\Omega_\text{S})\mathcal{D}(G_1)d\Omega_\text{S}dG_1,
\end{equation}

\noindent where $\chi$ is defined by Eq. (\ref{eq:single_dispersive_shift}) and 
$\langle\sigma_z\rangle(t)=\left(\langle\sigma_z\rangle(t_0) + 
1\right)e^{-t/\tilde{T}_1}$ is the spin polarization that results from a square pump pulse of duration 
$t_0$ and that has relaxed during a time lapse $t$ with a relaxation rate defined by 
Eq. (\ref{eq:ybridSpinRelaxationRate}). Here, the intrinsics relaxation time $T_{1} = 20$ s 
and decoherence time $T_2 = 200$ ns were taken
from experiments (Figs. \ref{fig:EPR}b and \ref{fig:shift-relaxation-1}b) and $\langle\sigma_z\rangle(t_0)$ has 
been computed from numerical simulations (detailed in the next section) 
that mimic the response to the driving pulse. The spin frequency distribution 
$\rho(\Omega_\text{S})$ (Fig. \ref{fig:shift-relaxation-2}b), arising from the inhomogeneous 
broadening, has been obtained from experimental results (Fig.  \ref{fig:spectroscopy}), whereas the 
spin-photon coupling distribution $\mathcal{D}(G_1)$ (Fig. \ref{fig:shift-relaxation-2}c) has been 
approximated to a $\sim G_{i}^{-3}$ power law dictated by the inductor geometry
(Fig. \ref{fig:shift-relaxation-2}d). Both distributions 
are considered to be mutually uncorrelated. This leaves the maximum spin-photon coupling 
as the only free parameter. Results obtained for two different magnetic fields, thus different average 
$\Delta$, are shown as solid lines in Fig. \ref{fig:shift-relaxation-2}a. The best fit provides a 
maximum $G_{1}$ as high as $100$ kHz for $N\sim10^2$ spins,
which we associate with those spins lying very close to the 
$w=50$ nm wide nanoconstriction fabricated at the inductor center.

%%%%%%%%%%%%% fig 18
\begin{figure*}[t!]
\begin{centering}
\includegraphics[width=1.98\columnwidth, angle=-0]{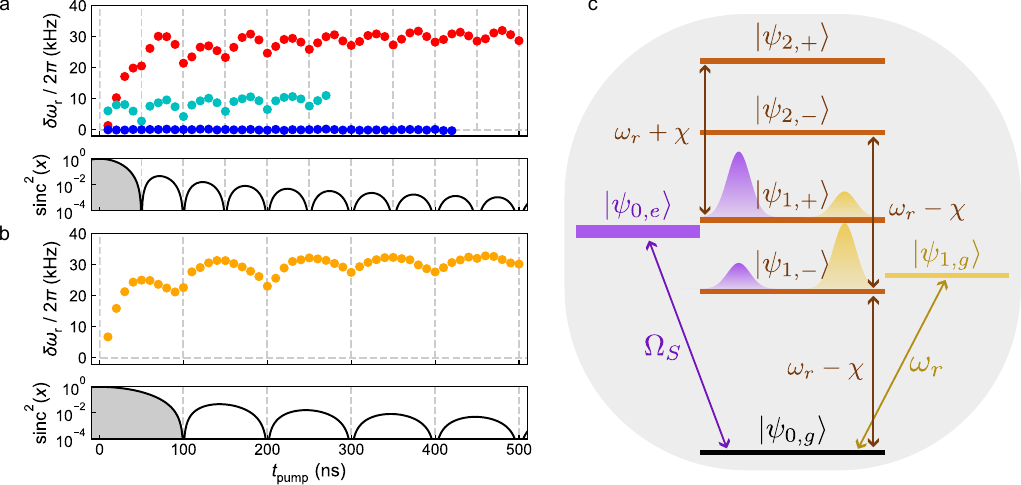}
\caption{(a) Top panel: Resonance frequency shift  
of a $2.734$ GHz LER generated by the application of 
square-shaped pump pulses with varying duration $t_{\rm pump}$ and resonant with the average 
PTMr spin frequency $\Omega_{\rm S}/2 \pi \simeq 2.754$ GHz at 
$B=98.36$ mT. The average spin-photon 
detuning $\Delta / 2 \pi \simeq 20$ MHz. The LER was 
coupled to $N \sim 5.5 \times 10^{11}$ PTMr spins and to independent readout and excitation 
transmission lines (Fig. \ref{fig:setup}). Bottom panel: square of the 
pulse Fourier transform amplitude at $\omega_{\rm r}$ as a function of pulse duration. Here, $x=(\omega_{\rm r}-\omega_{\rm pump})t_{\rm pump}/2$ (b) Same 
as in (a) for $\omega_{\rm pump} = \omega_{\rm r} + \Delta / 2$. Both measurements were performed 
at $T = 10$~mK. (c) Level scheme 
showing two spin excitation paths: direct resonant excitation of the spins by the 
main frequency component of the pump pulse and excitation of the LER by the sideband
pulse components.}
\label{fig:oscillations1}
\end{centering}
\end{figure*}

This value exceeds the maximum coupling $G_{1} \sim 32$ kHz estimated 
by extrapolating finite-element numerical simulations to spins located $\sim 1$ nm from the 
inductor surface (Fig. \ref{fig:frec before after fib and fields}d). The reason behind this discrepancy is not 
fully clear to us. It might be associated 
with the limitations of these numerical codes to simulate with sufficient resolution 
very small nanostructures with dimensions approaching the superconducting coherence 
length. Besides, the coupling between spins located very close to the Nb surface and 
the superconducting electrons might become of a different nature and stronger 
than that given by simple magnetostatic calculations, as has been revealed 
by experiments performed on individual molecules grafted on superconducting 
substrates \cite{Malavolti2018}. In either case, the results confirm that 
the combination of a suitable circuit design (low-$L$ LER) with local 
modifications via nanofabrication tools provides a 
promising method to approach the threshold for strong 
single spin to single photon 
coupling.

\subsection{Towards coherent spin manipulations}
\label{ssec:Rabi}

In order to coherently manipulate the spins, excitation pulses must be made 
shorter or of the order of the spin decoherence time $T_2 \sim 200$ ns of 
the PTMr deposits (Fig. \ref{fig:EPR}c). This condition calls for 
the application of high-power pulses through independent pump lines, 
exploiting the device design shown in Fig. \ref{fig:setup}d. This section 
describes the results of experiments performed in the dispersive regime 
using different pulse shapes, which explore different spin excitations.

We first consider spin excitations generated by square 
microwave pulses. Experiments were performed on a 
$\omega_{\rm r}/2 \pi = 2.73$ GHz low-$L$ NbTiN LER  
(Fig. \ref{fig:setup}) hosting $\simeq 5.5 \times 10^{11}$ 
PTMr molecules on its inductor line. The magnetic field 
$B = 98.36$ mT sets the average spin-LER detuning 
$\Delta / 2 \pi \simeq  20$ MHz, thus 
larger than the collective spin-photon coupling 
$G_{N}/ 2 \pi \simeq 2$ MHz. The 
resonance shift $\delta\omega_{\rm r}$ was measured for 
$\omega_{\rm{pump}}/2 \pi = 2.754$ GHz, which 
approximately matches the average $\Omega_{S}$. In 
order to expand the accessible time scales, the 
excitation pulse was made of 
three concatenated square-shaped pulses, with 
durations $t_{\rm pump}$, $2t_{\rm pump}$ and 
$t_{\rm pump}$, with $10$ ns 
$\leq t_{\rm pump} \leq 500$ ns.

The results are shown as red dots in 
Fig. \ref{fig:oscillations1}a. On top of a rapid 
increase and saturation, which takes place for 
$t_{\rm pump} \lesssim 100$ ns, an oscillatory 
component with a $50$ ns period shows up. Three control experiments were 
performed in order to clarify the origin of these  
two contributions. Pumping with the `mirror' frequency 
$\omega_{\rm{pump}} = \omega_{\rm r} - \Delta$, thus as close to the 
LER frequency as the previous one but fully detuned from the spin 
system, yields oscillations with the same amplitude and period, but on a 
much smaller base level (cyan dots in Fig. \ref{fig:oscillations1}a). This 
shows that the oscillatory contribution is not associated with spin nutations resonantly induced by the pump pulse. The blue dots in 
Fig. \ref{fig:oscillations1}a show the result of measurements performed 
at zero magnetic field, when the spins and the LER are fully 
decoupled. In this case, the pulse has no effect on 
$\omega_{\rm r}$. Finally, Fig. \ref{fig:oscillations1}b shows data 
measured for $\omega_{\rm{pump}} = \omega_{\rm r} + \Delta/2$. In this 
case, the oscillation period doubles to $100$ ns.

These results suggest that the fast $\delta \omega_{\rm r}$ oscillations 
depend on the detuning between the 
pump pulse and the LER, not the spins, but also that 
they reflect some spin excitations. These 
two effects can be understood as consequences of the use of 
square-shaped pulses. The Fourier transform amplitude of a 
finite-length pulse at frequency $\omega_{\rm r}$ is 
given by the cardinal sine function 
${\rm sinc}(x) = \sin(x)/x$, with 
$x = (\omega_{\rm r} - \omega_{\rm pump}) t_{\rm pump}/ 2$. 
The zeros of this function are located at $x = m \pi$, that 
is, $(\omega_{\rm r} - \omega_{\rm pump}) t_{\rm pump} / 2 \pi = m$, with $m \neq 0$ an 
integer, as shown by the bottom panels of Fig. \ref{fig:oscillations1}a-b. This is 
precisely the condition that is met by the values of 
$t_{\rm{pump}}$ at which $\delta \omega_{\rm r}$ minima are observed. 
It then follows that a small fraction of the pulse power, 
with frequency $\omega_{\rm r}$, directly drives the LER 
except when this condition is met. This suggests that 
the two components arise from, respectively, (i) 
direct resonant spin drive and (ii) sideband drive of 
the cavity excitations. 

The reason why exciting the superconducting cavity 
still results in a finite frequency shift $\delta \omega_{\rm r}$ can be 
understood by considering the nature of the excitations in 
the spin-LER coupled system. This picture is illustrated by 
the level scheme shown in Fig. \ref{fig:oscillations1}c, 
which for simplicity considers only the lowest lying 
resonator states. For a positive detuning 
($\Delta > 0$), the first two excited states can be 
approximated by:
\begin{equation} \label{eq:QubitLike_and_ResonatorLike_States}
	\begin{split}
		\ket{\psi_{1,\,+}} \simeq & \, \bigg|\dfrac{G_1}{\Delta} \bigg| \ket{\psi_{1,\,\rm{g}}} +  \ket{\psi_{0,\,\rm{e}}} 
        \textrm{,} \\
		\ket{\psi_{1,\,-}} \simeq & \, \ket{\psi_{1,\,\rm{g}}} - \bigg| \dfrac{G_1}{\Delta} \bigg| \ket{\psi_{0,\,\rm{e}}} \textrm{.}
	\end{split}
\end{equation}
Where `g' and `e' refer to the spin ensemble being in its ground state or 
having a single spin excitation, respectively. When no pump pulse is 
applied, the coupled system shows a resonance that corresponds 
to the $\ket{\psi_{0,\,\rm{g}}} \leftrightarrow \ket{\psi_{1,\,-}}$ 
transition. The pump pulse introduces two 
parallel paths towards the state $\ket{\psi_{1,\,+}}$. First, 
via the resonant spin excitation to $\ket{\psi_{0,\,\rm{e}}}$, induced by 
the main carrier frequency $\omega_{\rm pump} = \Omega_{\rm S}$. And 
second, via the sideband components of the square pulse with frequency 
$\simeq \omega_{\rm r}$ that generate $\ket{\psi_{1,\,\rm{g}}}$ whenever 
$(\omega_{\rm r} - \omega_{\rm pump}) t_{\rm{pump}} / 2 \pi \neq m$. In 
both cases, the system gets a positive frequency shift, as the further 
excitation $\ket{\psi_{1,\,+}} \leftrightarrow \ket{\psi_{2,\,+}}$ lies at 
$\omega_{\rm r} + \chi$ (see Fig. \ref{fig:oscillations1}c). These arguments remain 
valid for higher excited LER states, provided that the average number of 
photon excitations $n \ll N$, as it is the case in these 
experiments. We associate the rapidly saturating contribution 
observed experimentally to the former 
`conventional' dispersive shift, arising from the direct control of the 
spin polarization $\langle \sigma_{z} \rangle$, while the oscillations 
superimposed onto it correspond to the indirect spin excitation brought 
about by the tiny spin-cavity hybridization. The oscillations then probe 
to what extent such hybridization modifies wavefunctions at the given 
magnetic field (and detuning $\Delta$), as illustrated by 
Fig. \ref{fig:oscillations1}c.

\begin{figure}[h!]
\begin{centering}
\includegraphics[width=0.95\columnwidth, angle=-0]{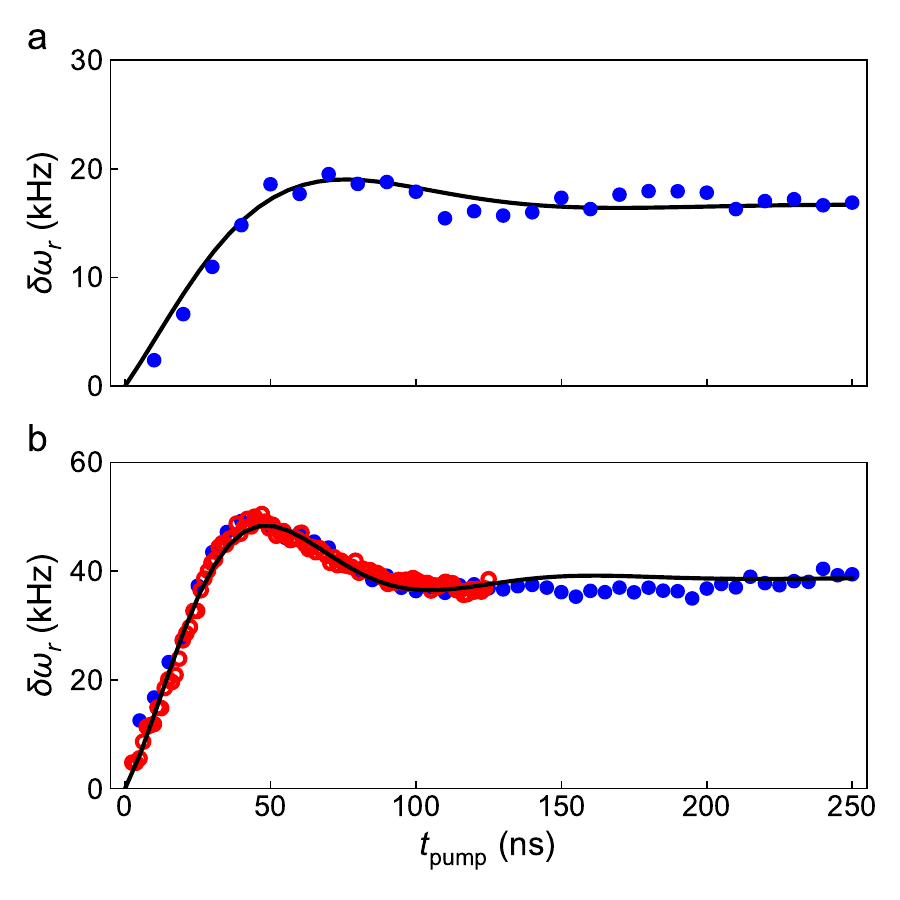}
\caption{(a) Frequency shift of a $2.734$ GHz LER associated with 
resonant spin excitations measured at $T=10$ mK as a function of the 
duration of a square-shaped pump pulse. The line is a least-squares fit with an 
exponentially damped oscillatory function 
$\delta \omega_{\rm r} (t_{\rm pump}) = \delta \omega_{\rm r}(0) e^{-(4t_{\rm pump}/\tau_{\rm R})} \sin{(4\Omega_{\rm R}t_{\rm pump})}$, which gives a Rabi frequency 
$\Omega_{\rm R}/2 \pi = 2.7 \pm 0.2$ MHz and a damping rate 
$1/2\pi \tau_{\rm R} = 5.8 \pm 0.6$ MHz. (b) Same as in (a) for a 
Gaussian shaped pump pulse, with $\Omega_{\rm R}/2 \pi = 4.44 \pm 0.04$ MHz 
and $1/2\pi \tau_{\rm R} = 6.61 \pm 0.14$ MHz.}
\label{fig:oscillations2}
\end{centering}
\end{figure}

Based on these qualitative considerations, 
subtracting the mirror-frequency ($\omega_{\rm r} - \Delta$) response cleanly 
isolates Rabi oscillations of the spin ensemble driven solely by the resonant 
excitation at $\omega_{\rm pump} = \Omega_{S}$.  
The result is shown in Fig. \ref{fig:oscillations2}a. The 
fit, which takes into account 
the total duration of the excitation pulse sequence, \emph{i.e.} 
$4t_{\rm pump}$, gives a Rabi frequency 
$\Omega_{\rm R} / 2 \pi = 2.7 \pm 0.2$ MHz and a decay time 
$\tau_{\rm R} = 172 \pm 18$ ns. The decay can be seen as a convolution of 
$T_{2}$ and of the inherent inhomogeneities of the system due to the 
distributions in qubit frequencies and in the coupling $G_{1}^{\rm line}$ 
of the spins to the driving transmission line.

A practical way to avoid driving the cavity is the use of Gaussian shaped 
pulses, which lack the sidebands of the ${\rm sinc}(x)$ function. 
Figure \ref{fig:oscillations2}b shows the result of repeating the 
time-dependent dispersive shift measurements using Gaussian 
excitation pulses. The pulse duration is defined here as 
$t_{\rm pump} = 2 \sigma$. The oscillation that was present 
for square pulses is gone, leaving only damped Rabi oscillations. A fit to 
an exponentially damped oscillatory function yields 
$\tau_{\rm R} = 151 \pm 3$ ns, and 
$\Omega_{\rm R} / 2 \pi = 4.44 \pm 0.04$ MHz.

%
%%%%%%%%%%%%% fig 19

Additional support to this interpretation can be found with the help of 
numerical simulations of the coupled spin-LER system dynamics. In 
order to make them as realistic as possible, they take into account all 
relevant effects mentioned in the preceding sections: inhomogeneous spin 
broadening $\gamma$, inhomogeneous spin-photon couplings $G_{1,j}$, the coupling to the 
transmission line $G_{1,j}^{\rm line}$, and the dissipation mechanisms of both the LER, at 
a rate $\kappa$, and of the spins, characterized by $1/T_1$ and $1/T_2$. 
Within the standard secular approximation, the response of the combined 
system is described by the system of equations
\begin{subequations}\label{eq:sim_EOM}
\begin{align}
    \frac{d}{dt}\langle a \rangle = &-i\tilde\omega_{\rm r}\langle a\rangle - \frac{i}{2}\sum_{j=1}^NG_{1,j}\langle S^-_j\rangle - i\sqrt{\kappa_c}\alpha_{\mathrm{in}}\ ,\label{eq:a_dynamic}\\ 
    \frac{d}{dt}\langle S^-_j \rangle = &-i\tilde\Omega_{S,j} \langle S^-_j\rangle + iG_{1,j}\langle S_j^z\rangle\langle a\rangle+ \notag\\
    & + iG_{1,j}^{\mathrm{line}}\langle S^z_j\rangle\ , \label{eq:sm_dynamic}\\
    \frac{d}{dt}\langle S^z_j \rangle  = &\frac{i}{2}G_{1,j}\left( \langle S_j^+\rangle\langle a\rangle - c.c. \right) - \frac{1}{T_{1}} \langle S^z_j\rangle- \notag\\
    & -\frac{i}{2}G_{1,j}^{\mathrm{line}}\left(\langle S^-_j\rangle - c.c.\right)\ ,\label{eq:sz_dynamic}
\end{align}
\end{subequations}
\noindent where $a$ and $S_{j}^{-}$ are photon and spin ladder operators, respectively, 
$\tilde\omega_{\rm r} = \omega_{\rm r} - \omega_\mathrm{pump} - i\kappa/2$ and $\tilde\Omega_{S,j} = \Omega_{S,j}-\omega_\mathrm{pump} - i/T_{2}$. 
Temperature effects are not included since 
the experiments described in this section have been carried out at constant 
$T$ such that $\hbar\Omega_S \gg k_\mathrm{B}T$. In Eq. (\ref{eq:a_dynamic}) we set 
$\alpha_{\mathrm{in}}^2 = P_{\mathrm{MW}}/\hbar\omega_\mathrm{pump}$, 
where $P_{\rm MW}$ is the microwave power fed into the pump line. In Eqs. (\ref{eq:sm_dynamic}) and (\ref{eq:sz_dynamic}), the coupling 
between the $j$-th spin and the transmission line, separated by a distance  $r_j$, 
is given by the classical magnetic field generated by the latter, 
$\hbar G_{1,j}^{\mathrm{line}} 
= g\mu_\mathrm{B} b_\mathrm{rms}^{\mathrm{line}}(r_j)$.

Equations (\ref{eq:sim_EOM}) have been solved using standard Runge-Kutta 
methods under certain approximations. Since the experimental sample 
comprises $\gtrsim 10^{11}$ spins, we are forced to reduce the number of 
equations by discretizing the distributions in spin frequencies $\rho(\Omega_{S})$ and spin-
photon couplings $\mathcal{D}(G_{1})$. The former is discretized into  $N_{\Omega}$ 
``boxes'' with $N_I$ spins in each of them, all having the same $\Omega_{S,I}$, so that 
$\sum_{I=1}^{N_{\Omega}} N_I = N$. Here, $N_I$ is  
proportional to the area encompassed by each ``box'' in the Gaussian 
distribution (Figs. \ref{fig:spectroscopy} and 
\ref{fig:shift-relaxation-2}b). On the other hand, both the coupling 
strengths $\{G_{1, J}\}_{J=1}^{N_{G}}$ and the number 
of spins $N_J$ having a coupling $G_{1,J}$ are estimated 
considering the inhomogeneity of the electromagnetic mode generated by the 
LER as determined in the previous section (Fig. \ref{fig:shift-relaxation-2}c 
and d). Since both distributions are uncorrelated, we consider that for all spin 
frequencies $\{\Omega_{S,I}\}_{I=1}^{N_{\Omega}}$, the distribution in 
couplings is the same. With these approximations, the number of equations 
in \eqref{eq:sim_EOM} is reduced from $2N+1 \sim 10^{12}$ to 
$2N_{\Omega}N_{G}+1 \ll N$.

\begin{figure}[h!]
\begin{centering}
\includegraphics[width=1\columnwidth, angle=-0]{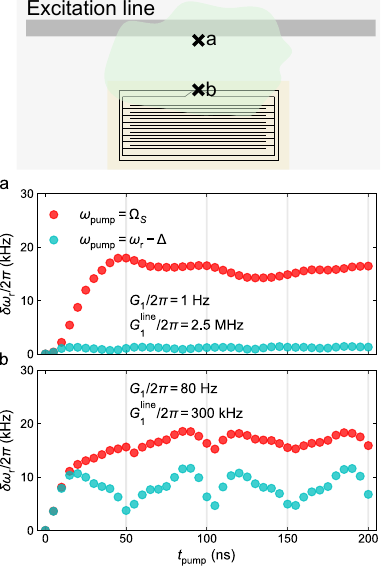}
\caption{Numerical simulations, by means of Eq. (\ref{eq:sz_dynamic}), of the frequency shift 
induced by the coupling of a LER to the two spin species `a' and `b' shown in the top scheme. 
(a) Those spins that are near the excitation line interact weakly with the LER photons. 
Hence, the population inversion 
occurs only when the driving frequency $\omega_{\rm pump}$ matches the spin resonant 
frequency $\Omega_{S}$. (b) In turn, those spins that are close to the inductor line, 
have a stronger interaction with the cavity at the expense of a smaller coupling to the 
excitation line. The population inversion in this case comes from the spin 
component in the hybridized wavefunction of the cavity photon. In both cases, a 
collection of $N_{\Omega}=51$ spin frequencies was considered and an arbitrary 
collective coupling $G_N = 0.6$ MHz was chosen
in order to make the amplitude of the signals comparable to those in
Fig.~\ref{fig:oscillations1}.}
\label{fig:dispersive_sim}
\end{centering}
\end{figure}

This analysis enables the identification of distinct 
contributions to $\delta \omega_{\rm r}$. We resolve two primary spin 
subsets: (a) spins strongly coupled to the excitation line and located far 
from the LER inductor (low $G_1$, high $G_1^{\text{line}}$), and 
(b) spins more strongly coupled to the LER photons, but less exposed to the 
excitation line (higher $G_1$, lower $G_1^{\text{line}}$). By applying 
Eq.~\eqref{eq:ensemble_dispersive_shift} to the numerically integrated 
trajectories from Eqs. \eqref{eq:sim_EOM}a–c, we obtain the dispersive 
shift generated by these two spin species.

Figure \ref{fig:dispersive_sim} shows illustrative results that 
follow from these simulations. Spins located near the excitation line 
(subset `a', Fig. \ref{fig:dispersive_sim}a) exhibit a 
damped Rabi oscillation when driven at $\omega_{\rm pump} = \Omega_S$ 
($=\omega_{\rm r}+\Delta$), but produce negligible signal under excitation at 
$\omega_{\rm pump} = \omega_{\rm r} - \Delta$. 
The Rabi frequency that characterizes the oscillation is then 
$\Omega_{\rm R}/2 \pi \simeq G_1^{\text{line}}/2 \pi \simeq 2.5$ MHz, 
which agrees very well 
with the experimental value (Fig. \ref{fig:oscillations2}). On the other 
hand, spins with higher coupling to the LER inductor 
(subset `b', Fig. \ref{fig:dispersive_sim}b) present the 
additional oscillatory response 
attributed to square pulse  
sidebands, which result from the spin-photon hybridization in the LER mode 
that is excited by these components. 

\section{Conclusions}
\label{sec:conclusions}
We have demonstrated tunable strong coupling between 
molecular spin ensembles and superconducting lumped-element resonators. 
Besides, we have experimentally 
shown the crucial role played by the coupling to the 
readout line in order to achieve a good  
visibility of the two spin-photon polariton branches 
in microwave transmission.

In these experiments, however, the strong spin-photon 
coupling $G_{N}$ arises mainly from the relatively high 
number of spins ($\gtrsim 10^{12}$) that are on 
`speaking terms' with the LER inductor. Its collective 
nature has been established by modifying both temperature 
and the spin concentration in the molecular ensemble. The 
average single spin coupling $G_{1}/2 \pi \simeq 6$ Hz is 
still modest, although somewhat larger than the typical 
values obtained with coplanar resonators, and remains far 
below the decoherence rates of the molecular spins. 
Reducing the inductor to a single wire 
provides a simple method to further enhance $G_{1}$ by up 
to a factor $2$, even at the cost of decreasing $G_{N}$.

The same platform allows studying the molecular spin 
relaxation and its coherent dynamics, using excitation and 
detection stages that are fully integrated in the same 
chip. Pump-probe experiments performed in the dispersive 
regime, when spins are energetically detuned from 
the LER, show the ability to detect LER frequency 
shifts associated with changes in the spin states. 
The spin polarization determined in this way as a function 
of temperature, shows that one can reach a close to full 
spin initialization for temperatures approaching 
$10-20$ mK and magnetic fields $B \gtrsim 0.1$ T (or, 
equivalently, spin frequencies 
$\Omega_{S}/2 \pi \gtrsim 2.5$ GHz). In addition, it 
provides a method to determine the spectrum of spin 
excitations.

Of especial interest are the results of relaxation 
experiments. We find that the coupling of the spins to 
cavity photons can provide a fast path for 
the spin initialization through the Purcell effect. The 
results have allowed a quantitative determination of the 
distribution of single-spin couplings, and confirm the 
very large enhancement of $G_{1}$ for those spins located 
near a nanoconstriction fabricated in the inductor line. 
We have estimated $G_{1}$ values as high as $100$ kHz, 
which to our knowledge are the maximum reported to this 
date.

The introduction of short and intense microwave pulses 
through an independent line demonstrates  
coherent control of molecular spins within a fully integrated 
superconducting circuit, closing the loop from initialization to manipulation and 
readout. The experiments, backed by numerical simulations of the 
spin-LER coupled dynamics, show that the spins can be 
either excited directly by the pump line or indirectly via 
sideband excitations of LER modes. In the former case, 
the pulse frequency can be varied freely to address 
specific parts of the spin spectrum. In the latter case, 
the results provide a measure of the hybridization between 
spins and cavity photons. The relative 
intensities of these two processes depends on the location 
of the spins with respect to the inductor and to the 
pumping line, and can be tuned by shaping the pumping 
pulses. This dependence introduces also a limitation for 
the implementation of coherent control on spin ensembles, 
since each spin is driven by a different microwave 
field. This could be improved by designing pumping lines 
focusing the field near the same sample regions, e.g. near 
a nano-constriction, which couple most to the LER, thus 
combining a maximum drive and higher Rabi frequencies with 
maximum detection sensitivity. This can be complemented by 
adequately shaping the pulses, e.g. using the BB1 
sequence, which can mitigate the error generated by a small microwave field inhomogeneity \cite{Morton2005}.

The results of this work show that lumped element 
resonators coupled to independent readout and pumping 
lines incorporate most of the ingredients, i.e. 
initialization, control and detection, needed to implement 
a quantum computation architecture with molecular 
spins \cite{Chiesa2023}. In particular, the combination of 
sufficiently low inductance with nanoscopic constrictions 
helps reaching locally single spin to photon coupling 
strengths that can overcome the molecular spin 
decoherence rates. The situation could be improved even 
further with the use of LERs with parallel plate 
capacitors, which eliminate parasitic inductance present 
in finger capacitors \cite{Chiesa2023}. Besides, the high 
inhomogeneity in the spin-photon interaction could be 
advantageous in order to exploit circuit `hot spots' for 
control and readout of specific spins, even when dealing 
with a large ensemble. This idea can be extended to 
molecules with a richer spin level spectrum, i.e. qudits, 
since the pumping line admits broadband pulses, bringing the implementation of quantum 
simulation \cite{Roca2025simulating} and quantum error correction \cite{Chiesa2020} algorithms closer to this type 
of platform. Future work will combine optimized circuit geometries with 
chemically engineered spin isolation and organization, 
exploiting techniques such as self-assembly \cite{Tesi2023} 
or Langmuir-Blodgett \cite{Gimeno2025} deposition, to 
realize coherent operations on molecular spin qubits and 
qudits assembled on a chip. 

%%%%%%%%%% acknowledgments

\begin{acknowledgments}
L.T. acknowledges Prof. Thomas Prisner and the 
Biomolekulares Magnetresonanz Zentrum (BMRZ) for providing 
access to the pulsed EPR spectrometer, and Dr. Burkhard 
Endeward for his support with the pulsed EPR setup. This 
work has received support from grants 
TED2021-131447B-C21, TED2021-131447B-C22, 
PID2022-140923NB-C21, PID2022-137779OB-C41, PID2022-137779OB-C42, PID2022-137779OB-C43, 
PID2022-141393OB-I0, PID2022-137332OB-I00, 
PID2022-137779OB-C41, CEX2020-001039-S, CEX2023-001263-S, and CEX2023-001286-S, funded by 
MCIN/AEI/10.13039/501100011033, 
ERDF `A way of making Europe' and ESF `Investing in 
your future', from Deutsche Forschungsgemeinschaft with the 
Emmy Noether project number 529038510, from the Gobierno de 
Arag\'on grant E09-23R-QMAD and from Generalitat de 
Catalunya grants SGR Cat 2021-00438 and 2021-SGR-00443. We 
also acknowledge funding from the 
European Union Horizon 2020 research and innovation 
programme through FET-OPEN grant FATMOLS-No862893 and from 
QUANTERA project OpTriBits (AEI: PC2024-153480 and DFG 
532763805). This study forms also part of the Advanced 
Materials and Quantum Communication programmes, 
supported by MCIN with funding from European Union 
NextGenerationEU (PRTR-C17.I1), by Gobierno de 
Arag\'on, and by CSIC (QTEP-PT-001).
\end{acknowledgments}

\newpage
%%%%%%%%%% bibliography

\bibliographystyle{apsrev4-1}
\bibliography{PTMr}

\end{document}